\documentclass[sigconf,screen]{acmart}

\usepackage[normalem]{ulem}

\usepackage{fancyhdr}
\usepackage[normalem]{ulem}

\usepackage{listings}
\usepackage{amsmath}
\usepackage[noend,linesnumbered,ruled,vlined]{algorithm2e}
\usepackage{physics}
\usepackage{amsmath}
\usepackage{multirow}
\usepackage{enumitem}

\usepackage{amssymb}
\newcommand{\ignore}[1]{}
\usepackage{adjustbox}
\usepackage[normalem]{ulem}
\usepackage[flushleft]{threeparttable}
\usepackage{dblfloatfix}

\usepackage{tikz}
\usetikzlibrary{quantikz}
\usepackage{subcaption}

\usepackage{color,soul}
\usepackage{url}

\def\BibTeX{{\rm B\kern-.05em{\sc i\kern-.025em b}\kern-.08em
    T\kern-.1667em\lower.7ex\hbox{E}\kern-.125emX}}
\definecolor{aliceblue}{rgb}{0.93, 0.9, 0.95}
\usepackage{xcolor}
\usepackage{tcolorbox}
\tcbset{width=1\columnwidth,boxrule=1pt,colback=aliceblue,arc=1.5pt,left=2pt,right=2pt,boxsep=0.5pt}

\definecolor{Burgundy1}{RGB}{128,0,32}

\newcommand\mq[1]{%
  \bgroup
  \hskip0pt\color{Burgundy1}%
  #1%
  \egroup
}

\usepackage{tikz}

\def\BibTeX{{\rm B\kern-.05em{\sc i\kern-.025em b}\kern-.08em
    T\kern-.1667em\lower.7ex\hbox{E}\kern-.125emX}}
    
\def\BibTeX{{\rm B\kern-.05em{\sc i\kern-.025em b}\kern-.08em
    T\kern-.1667em\lower.7ex\hbox{E}\kern-.125emX}}

\usepackage{balance}

\AtBeginDocument{%
  \providecommand\BibTeX{{%
    \normalfont B\kern-0.5em{\scshape i\kern-0.25em b}\kern-0.8em\TeX}}}

\setcopyright{none}
\acmDOI{10.1145/3503222.3507703}
\acmYear{2022}
\copyrightyear{2022}
\acmSubmissionID{asplos22main-p56-p}
\acmISBN{978-1-4503-9205-1/22/02}
\acmConference[ASPLOS '22]{Proceedings of the 27th ACM International Conference on Architectural Support for Programming Languages and Operating Systems}{February 28 -- March 4, 2022}{Lausanne, Switzerland}
\acmBooktitle{Proceedings of the 27th ACM International Conference on Architectural Support for Programming Languages and Operating Systems (ASPLOS '22), February 28 -- March 4, 2022, Lausanne, Switzerland}

\begin{document}

\title[HAMMER: Boosting Fidelity of Noisy Quantum Circuits by Exploiting Hamming Behavior ...]{HAMMER: Boosting Fidelity of Noisy Quantum Circuits\\by Exploiting Hamming Behavior of Erroneous Outcomes}


\settopmatter{authorsperrow=2}

\author{Swamit Tannu$^*$}
\affiliation{%
  \institution{University of Wisconsin–Madison}
  \city{Madison}
  \country{USA}  
  }    
  
\author{Poulami Das}
\affiliation{%
  \institution{Georgia Institute of Technology}
  \city{Atlanta}
  \country{USA}
}

\author{Ramin Ayanzadeh}
\affiliation{%
\institution{Georgia Institute of Technology}
  \city{Atlanta}
  \country{USA}
}
\author{Moinuddin Qureshi}
\affiliation{%
\institution{Georgia Institute of Technology}
  \city{Atlanta}
  \country{USA}
}

\renewcommand{\shortauthors}{Swamit Tannu, Poulami Das, Ramin Ayanzadeh, Moinuddin Qureshi}


\vspace{0.2 in}
\begin{abstract}

Quantum computers with hundreds of qubits will be available soon. Unfortunately, high  device error-rates pose a significant challenge in using these near-term quantum systems to power real-world applications. Executing a program on existing quantum systems generates both correct and incorrect outcomes, but often, the output distribution is too noisy to distinguish between them.
In this paper, we show that erroneous outcomes are not arbitrary but exhibit a well-defined structure when represented in the Hamming space. 
Our experiments on IBM and Google quantum computers show that the most frequent erroneous outcomes are more likely to be close in the Hamming space to the correct outcome. We exploit this behavior to improve the ability to infer the correct outcome.

We propose {\em Hamming Reconstruction (HAMMER)}, a post- processing technique that leverages the observation of Hamming behavior to reconstruct the noisy output distribution, such that the resulting distribution has higher fidelity. We evaluate HAMMER using experimental data from Google and IBM quantum computers with more than 500 unique quantum circuits and obtain an average improvement of 1.37x in the quality of solution. On Google's publicly available QAOA datasets, we show that HAMMER sharpens the gradients on the cost function landscape.

\end{abstract}


\maketitle

\pagestyle{plain}

\section{Introduction}

Quantum computers with few tens of quantum bits (qubits) are currently available to users, and machines with several hundred qubits are expected within the next few years~\cite{ibmq1000qubitroadmap}. These machines can accelerate certain optimization problems~\cite{qaoa}, molecular simulations~\cite{vqe}, and machine learning~\cite{biamonte2017quantum} applications. While their prospective looks promising, near-term quantum computers are noisy and prone to device errors, which severely limits the fidelity of applications. Although the number of qubits on quantum computers has notably increased in recent years, the average error rate of quantum operations has not reduced at the same pace. For example, the average two-qubit operational error rate on existing quantum computers from IBM and Google continues to be in the range of 1\% to 2\%~\cite{IBMQ,sycamoredatasheet}. Such large error rates limit the number of operations that can be performed reliably. 
As quantum machines are unlikely to have enough resources to support error correction in the near term, such machines are typically operated in the  {\em Noisy Intermediate-Scale Quantum (NISQ)}~\cite{preskillNISQ} mode, whereby the program is executed several thousand times, and each trial can produce a correct or an erroneous outcome. The application fidelity on a NISQ machine is determined by the ability to identify the correct answer from the outcomes produced during all the trials.

\begin{figure*}[htp]
\centering
    \vspace{0.15 in}
    \includegraphics[width=0.90\textwidth]{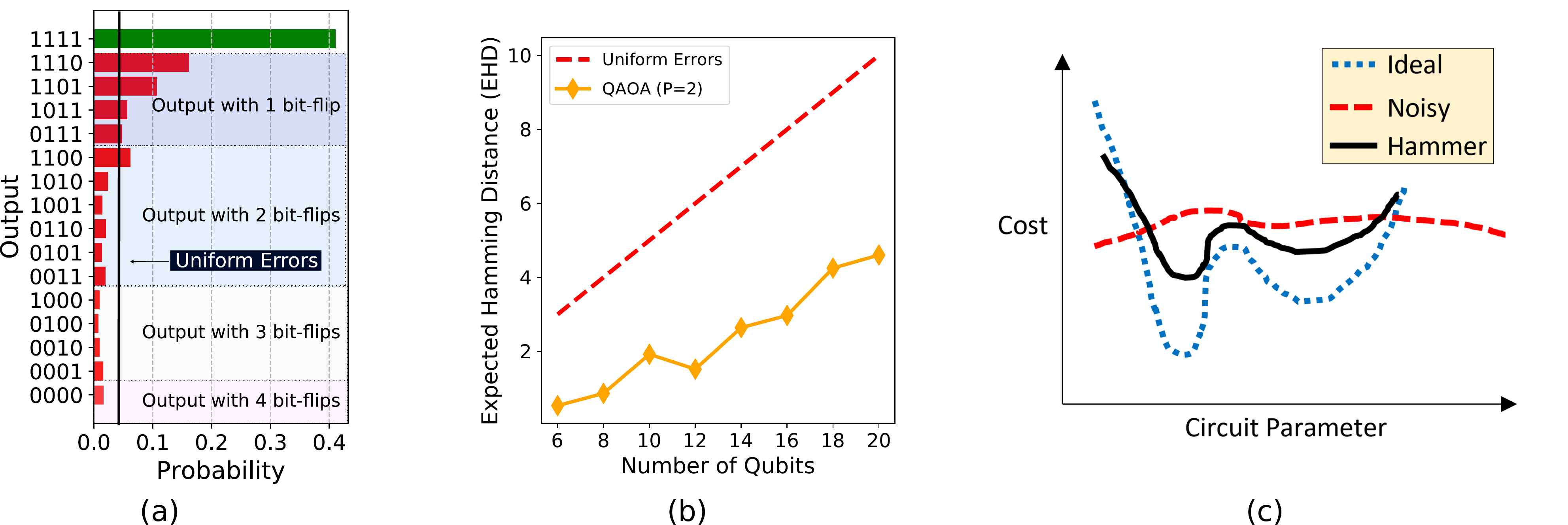}
    \vspace{-0.1 in}
    \caption{(a) Histogram of the output distribution for a 4-qubit Bernstein-Vazirani circuit. (b) Trend in the Expected Hamming Distance in the output distribution for \texttt{QAOA} circuits, run on IBM-Paris. (c) Cost landscape of a Variational Quantum Circuit.} 
    \vspace{0.2 in}
    \label{fig:introfigure}
\end{figure*}

To improve the fidelity of applications on NISQ hardware, recent works propose compiler techniques that reduce the number of quantum operations (gates)~\cite{nishio,li2018tackling,shi2019optimized,shi2020resource,alam2020circuit,zulehner2018efficient}, 
perform error-aware computations~\cite{noiseadaptive,tannu2019not,micro3}, focus on reducing specific sources of errors, such as measurement errors~\cite{kwon2020hybrid,matrixmeasurementmitigation,funcke2020measurement,FNM,das2021jigsaw}, idling errors~\cite{smith2021error,das2021adapt}, and crosstalk~\cite{murali2019noise}. Recent works~\cite{micro1,patel2020veritas} have also looked at the problem of mitigating correlated errors, where a particular incorrect outcome can occur with a high probability. The implicit assumption in all these prior approaches is that the erroneous outcomes do not provide any meaningful information.

Even with intelligent compilation techniques, NISQ machines produce incorrect or sub-optimal outcomes for a significant fraction of the trials. In this paper, we propose an orthogonal approach to improve the fidelity of NISQ programs. Rather than simply treating the erroneous outcomes as wasteful, we propose to leverage correlation in such erroneous outcomes. In particular, we try to address the following two questions in this paper:
\begin{enumerate}
    \item When a trial produces a wrong outcome, is the produced bitstring arbitrary, or does it have a specific structure?
    \item If erroneous outcomes have some structure, can we exploit that structure to improve the quality of the solution?
\end{enumerate}

To understand the behavior of the incorrect outcomes produced during program execution, we analyze experimental data from three different IBM quantum computers and publicly available datasets from Google with more than 500 representative circuits containing up to 20 qubits. If the incorrect outcomes did not have any structure, we would expect all possible incorrect outcomes to occur with close to uniform probability. However, the incorrect outcomes have a well-defined structure in the Hamming space, as the incorrect outcomes tend to be within a short Hamming distance from the correct or optimal answers. For example, Figure~\ref{fig:introfigure}(a) shows the output histogram of a four-qubit Bernstein-Vazirani (\texttt{BV}) circuit in which the error-free output "1111" appears with only 40\% probability. We observe that the most frequent incorrect outcomes are close to the correct output in the Hamming space. 

We observe a similar structure in the output of variational quantum programs such as {\em Quantum Approximate Optimization Algorithm (\texttt{QAOA})}. To understand the structure of errors in Hamming space, we compute the {\em Expected Hamming Distance (EHD)}, which is a weighted average of the Hamming distances between correct and incorrect outcomes. EHD captures the density of outcomes in the Hamming space. If the erroneous outcomes produced are arbitrary for an n-qubit program, then we would expect that the incorrect bitstring will be $\frac{n}{2}$ bits away from the correct outcomes on an average. Alternately, when outcomes with high probabilities are clustered around the correct answers, the EHD would approach to zero. Figure~\ref{fig:introfigure}(b) shows the EHD for \texttt{QAOA} circuits with two layers ($p=2$)~\cite{qaoa} as the size of the circuit is increased from 5 qubits to 20 qubits. \ignore{Note that `$p$' in \texttt{QAOA} denotes a $p$-layer parametric quantum circuit with $2p$ variational parameters, and in theory, the performance of \texttt{QAOA} is expected to increase with an increasing value of $p$. The depth of these circuits increases linearly with the number of qubits.}  We observe that although the EHD increases with the number of qubits, it increases at a much slower pace compared to a uniform-error model. Thus, incorrect outcomes are not arbitrary but exhibit a well-defined structure in the Hamming space.

\begin{figure*}[t]
\centering
    \includegraphics[width=2\columnwidth]{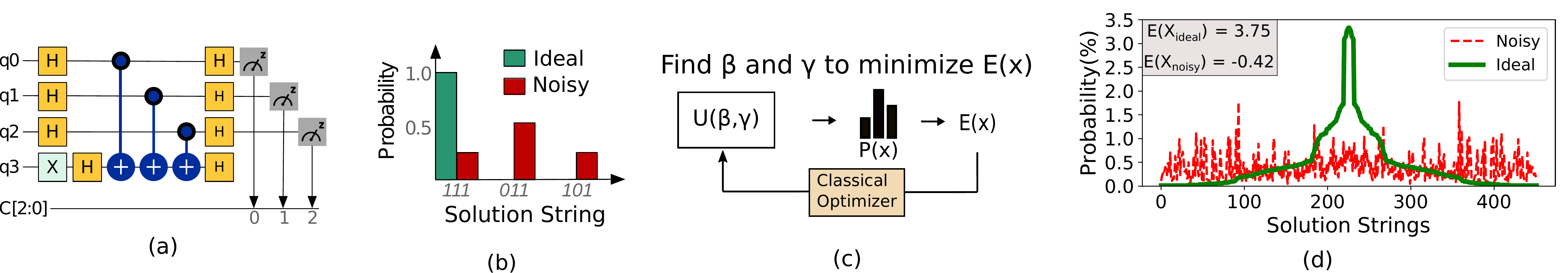}
    \caption{(a) Example of a 3-qubit Bernstein-Vazirani circuit. (b) Output on ideal (noise-free) and noisy NISQ hardware. (c) Variational Quantum Approximate Optimization Algorithm (\texttt{QAOA}). (d) Ideal and real output of a \texttt{QAOA-9} circuit on IBM-Paris.}
    \vspace{0.1 in}
    \label{fig:bvcircuitexample}
\end{figure*}

As erroneous outcomes are not arbitrary, it is possible to leverage the structure they exhibit in the Hamming space. To this end, we propose {\em Hamming Reconstruction ({\em HAMMER})}, a post-processing technique for boosting the fidelity of NISQ applications. Instead of relying solely on the probabilities associated with each outcome, HAMMER analyzes their neighborhoods in the Hamming space. More specifically, by utilizing the probabilities of individual outcomes and their structure in Hamming space, HAMMER provides more accurate estimates of the \textit{likelihood} of each outcome. By using this likelihood function, HAMMER boosts the outcomes that are likely to be correct, while "hammering" down those that are potentially incorrect.

We observe that despite the probability of obtaining the correct bitstring being low for large quantum circuits, the correct answer has a rich neighborhood, HAMMER leverage this insight to compute a Neighborhood Score for every outcome string. The neighborhood Score is computed by obtaining a weighted sum of all the strings that are $k$ Hamming distance away. \ignore{ For example, the Hamming spectrum of “1111” (the correct output), shown in Figure \ref{fig:introfigure}(a),  consists of five buckets corresponding to Hamming distances ranging from 0-4. Here, Bucket-0 includes the only outcome at Hamming distance zero with “1111” (which is "1111" itself), Bucket-1 includes all one Hamming distance away outcomes, and so on. In the next step, HAMMER uses the probabilities of the entries in every bucket to calculate the Hamming score (HS) of the bucket for the corresponding outcome (here "1111"). However, note that the knowledge of correct vs. incorrect outcomes are unknown.} Therefore, HAMMER uses a consensus from the individual Hamming scores of all the outcomes and generates a \textit{weight} for evaluating the neighborhood at each Hamming distance. By aggregating these weights, HAMMER determines the final Neighbourhood Score for each outcome that is used in conjunction with its probability to estimate the likelihood of the outcome. As HAMMER uses the knowledge of the neighborhood, it boosts the probabilities of the correct outcomes while diminishing the probabilities of the incorrect ones. 

Not only can HAMMER improve the fidelity of applications that produce a single correct output, but also of near-term variational quantum algorithms. For example, Figure~\ref{fig:introfigure}(c) shows the landscape of the cost function of a \texttt{QAOA} circuit, as the circuit parameters are tuned. Unfortunately, due to high error rates, a significant fraction of the outcomes in the output distribution are sub-optimal. Therefore, the expected cost becomes insensitive to changes in the circuit parameters, which impedes the search for the optimal solution. As HAMMER builds the consensus using Hamming Spectrum, it amplifies the probabilities of the outcomes with Hamming structure and attenuates the spurious solutions to improve the average quality of solutions, thus helping the search for high-quality solutions. 

\vspace{0.1 in}

Overall, our paper makes the following contributions:

\begin{enumerate}

\item To the best of our knowledge, this is the first paper to demonstrate that incorrect outcomes produced on NISQ machines are not arbitrary but manifest a well-defined structure in the Hamming space. In particular, incorrect outcomes tend to be at a short Hamming distance from the correct outcome.

\item We propose HAMMER, a post-processing technique that exploits the Hamming structure of incorrect outcomes to boost the likelihood of the correct outcomes.  
\item We evaluate HAMMER on Google and IBM machines with more than 500 circuits and show that HAMMER provides a consistent improvement in the average quality of solutions.     
\end{enumerate}
\section{Background}

\subsection{NISQ Paradigm}

Existing qubit devices are vulnerable to noise. The average gate error rates range from 0.1\% to 2\% on current IBM and Google quantum computers. Such a high error rate limits the size of the largest quantum circuit that can be executed as the probability of encountering errors during computations increases with the number of operations. To run practical quantum applications, we need to protect against hardware errors. Unfortunately, quantum error correction codes incur large overheads requiring thousands of physical qubits. Thus, error correction on near-term systems with a few hundreds of qubits is not plausible. In the near-term, we have to operate quantum computers without error correction. Although error-prone, these {\em Noisy Intermediate Scale Quantum (NISQ)}~\cite{preskillNISQ} computers are still considered promising for very specific applications~\cite{qaoa,vqe}. Consequently, there is an active effort in devising algorithms, hardware (building better devices and gates), and software (compiler and post-processing) solutions to reduce the impact of errors.

\subsection{Impact of Noise on Quantum Circuits}
NISQ machines are vulnerable to errors and often produce incorrect outcomes. For example, Figure~\ref{fig:bvcircuitexample}(a) shows a 3-qubit Bernstein Vazirani (\texttt{BV}) circuit that encodes the secret key "\textit{111}". Ideally, this circuit should produce the output in a single query on a quantum computer. However, due to hardware errors, quantum computers produce incorrect outcomes "\textit{011}" and "\textit{101}" in addition to the correct outcome "\textit{111}". Executing a quantum program on noisy hardware produces a large number of incorrect outcomes, along with the correct ones. Such incorrect bitstrings may occur with a very high probability, and the output distribution can become too noisy that the correct outcome is indistinguishable from the incorrect ones.

\begin{figure*}[t]
\centering
    \includegraphics[width=2\columnwidth]{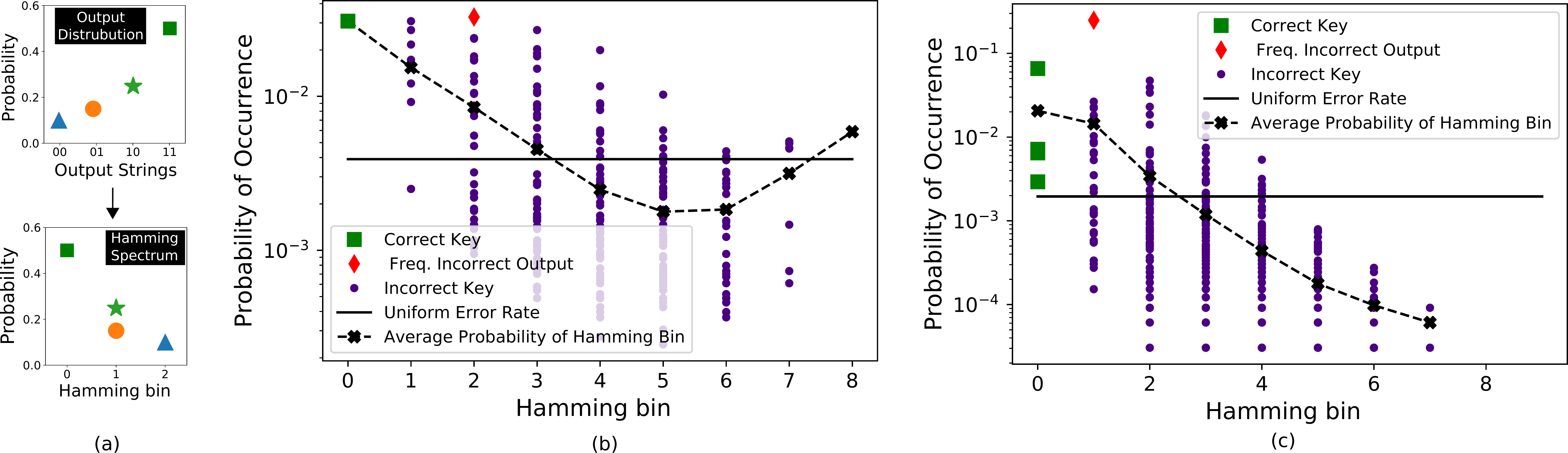}
    \vspace{0.05 in}
    \caption{(a) Representing output probability distribution as Hamming spectrum (illustrative example).  (b) Hamming Spectrum of a \texttt{BV-8} circuit (executed on IBM-Manhattan).  (c) Hamming Spectrum of a \texttt{QAOA-8} circuit (executed on IBM-Manhattan).}
    \label{fig:histogram}
\end{figure*}

\subsection {Variational Quantum Algorithms}
Despite the vulnerability of NISQ machines to hardware errors, we can still use these machines for solving a class of optimization problems using variational quantum algorithms (VQA), which are robust to a certain types of errors. VQAs use a parametric circuit and search iteratively for the circuit parameters that produce high-quality solutions. For example, when solving an optimization problem using \texttt{QAOA}, as illustrated in Figure~\ref{fig:bvcircuitexample}(c), we search for circuit parameters $\beta$ and $\gamma$ using a two-step process. First, we initialize $\beta$ and $\gamma$ with the best-known value and execute a quantum circuit several thousands of times. This yields a distribution of solution strings, where each solution has a fixed cost. Our objective is to find the solution string with the lowest cost. Next, we perform a second step, in which we compute the average (or expected) cost corresponding to the output distribution and search for optimal $\beta$ and $\gamma$ using expected cost as the objective function. 

Unfortunately, high error rates of NISQ hardware disrupt the variational loop as the output distributions can be extremely noisy. Figure~\ref{fig:bvcircuitexample}(d) shows an ideal distribution and the output on an IBMQ machine for a \texttt{QAOA-9} benchmark. Due to the high error rate, a significant fraction of outcomes are solutions with suboptimal costs. These suboptimal outcomes result in inaccurate estimation of the expected cost on NISQ devices. For example, ideally the expected cost should be E(x) = 3.75 but in reality, it is E(x)=-0.42, as shown in Figure~\ref{fig:bvcircuitexample}(d). Moreover, due to increasing noise, the expected cost becomes insensitive to changes in $\beta$ and $\gamma$ and the cost function landscape plateaus. This makes optimization problems at practical scales beyond the reach of \texttt{QAOA} on existing NISQ hardware.

\subsection{Improving Quality of Solution on NISQ}

Existing error-mitigation techniques focus on providing better than worst-case reliability on NISQ machines. The implicit assumption in these techniques is that erroneous outcomes may have no meaningful information to determine the correct answer. This would be true if the values produced by the erroneous trials were arbitrary -- with a uniform probability over all possible incorrect answers.  However, if the incorrect values have some correlation with the correct answer, then we can analyze the incorrect values to determine the correct answer.

\section{Hamming Behavior of Errors}
\label{sec:char}

\subsection{Is there a structure in Errors?}

To understand the structure in errors, we run a \texttt{GHZ-10} circuit. On an ideal (error-free) quantum computer, the output state is an equal superposition of the all-zero and all-one states. However, on the IBM quantum computer, hardware errors produce incorrect outcomes along with the two desired states. In the case of the \texttt{GHZ-10} circuit, we observe that correct outcomes occur with a cumulative probability of 45\%, while 55\% is the collective probability of all the incorrect outcomes. The incorrect outcomes are not random as the dominant incorrect outcomes that appear with high frequency are close to the correct answers in Hamming space. We also observe that majority of the dominant incorrect outcomes are within a Hamming distance of two from either correct answer.      

The Hamming behavior of erroneous outcomes is not unique to \texttt{GHZ}. We observe a similar pattern in errors by analyzing data from more than 1500 quantum circuits executed on IBM and Google quantum hardware. These circuits capture diverse trends in the number of gates, circuit depth, degree of entanglement, and measurement basis. A detailed analysis on how these factors impact the Hamming structure is presented in the Section~\ref{sec:entangle}.

\subsection{Hamming Spectrum} 

To visualize the structure in errors, we illustrate the output distribution in the form of a {\em Hamming spectrum}. Hamming spectrum creates a compact representation of the output probability distribution by bucketing each outcome into Hamming bins, as shown in Figure~\ref{fig:histogram}(a). Every string outcome in the output distribution is added to the $K^{th}$ bin, where $K$ is the Hamming distance between the correct answer(s) and the string. For a quantum circuit with an N-bit output, $K$ ranges between 0 to N. Figure~\ref{fig:histogram}(b) shows the Hamming spectrum of \texttt{BV-8} output, where correct output is the all-one state ("11111111"), and rest are erroneous outcomes. In the Hamming spectrum, we highlight - (1) correct output (2) an erroneous outcome that occurs more frequently than the correct output (3) rest of the erroneous outcomes (4) average of the Hamming bin. 

In the Hamming spectrum shown in Figure~\ref{fig:histogram}(b), we observe that many incorrect outcomes that appear with high probability are close to the correct answer in Hamming space. Furthermore, the probability of an output string in a given Hamming bin reduces with increasing Hamming distance. For example, incorrect outputs that are four Hamming distances away from the correct answer have a lower than a random chance of occurrence. Figure~\ref{fig:histogram}(b) also shows the uniform probability distribution where all $2^{N}$ outcomes are equally likely with $\frac{1}{2^{N}}$ probability. 
So far, we have used \texttt{BV} to illustrate the structure in the output errors. However, unlike \texttt{BV}, most practical NISQ circuits produce output distributions with multiple correct outcomes. To understand if the structure in errors persists for such circuits, we analyze representative \texttt{QAOA} circuits that produce multiple correct outcomes. Figure~\ref{fig:histogram}(c) shows the Hamming spectrum for a \texttt{QAOA-8} circuit that uses eight qubits and produces three correct outcomes with 82\%,10.5\% and 7.0\% probability, respectively, in the absence of noise. On IBM Manhattan, the three dominant solutions appear with 7\%, 0.7\%, and 0.2\% probability respectively. Most incorrect answers produced while running \texttt{QAOA} are within three Hamming distance from the correct answers. Note that for multiple correct answers, we consider the shortest Hamming distance.

\begin{figure*}[htp]
\centering
\vspace{0.1 in}
    \includegraphics[width=\textwidth]{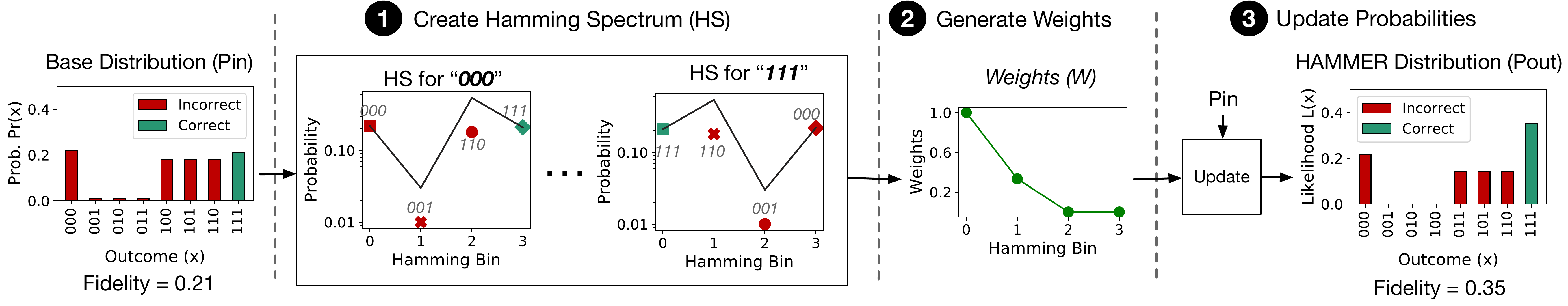}
    \caption{Overview of HAMMER to translate the distribution generated by the NISQ machine to corrected distribution.} 
    \vspace{0.1 in}
    \label{fig:hammerblockdiagram}
\end{figure*}

\subsection{How to Quantify Hamming Behavior?}

To quantify the degree of Hamming structure, we use {\em Expected Hamming Distance (EHD)}. EHD computes the weighted sum of Hamming distances between correct outcome(s) and incorrect observations, where the weights are the probabilities of the incorrect observations. 
For output distributions without errors, EHD is zero. Whereas, in case of a uniform output probability distribution, EHD approaches $\frac{n}{2}$, where n is the number of qubits. Note that EHD $\in$ [0,n] as the output strings can be up to "n" hamming distance away from the correct outcomes. A quantum circuit that uses n qubits and produces correct answers all the time will have an EHD of zero, whereas, for uniform output probability distribution where all outcomes are equally likely, EHD would be close to ~$\frac{n}{2}$.

\begin{figure}[b]
\centering
    \vspace{-0.1 in}
    \includegraphics[width=\columnwidth]{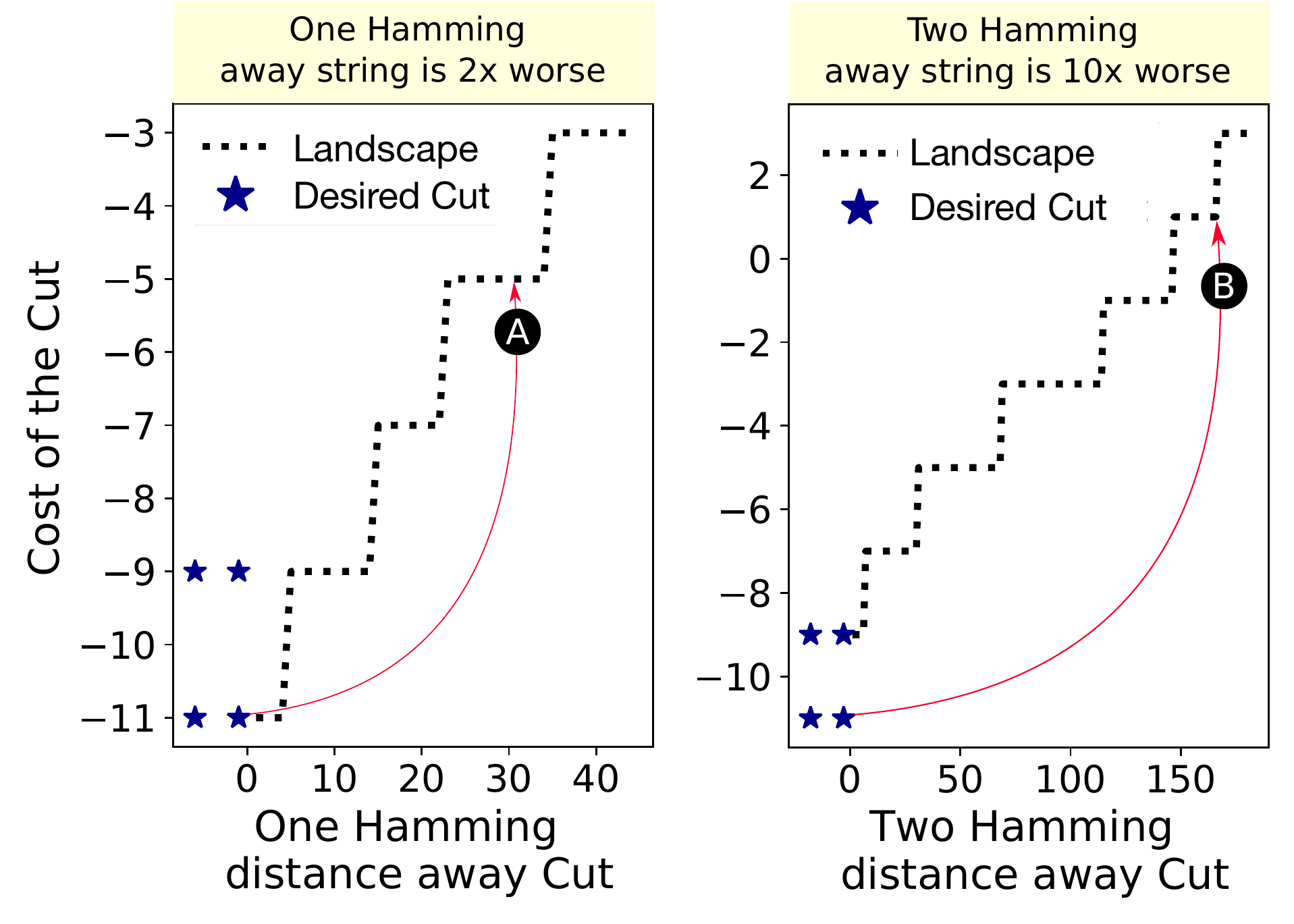} 
    \vspace{-0.05 in}
    \caption{Landscape of a \texttt{QAOA-10} benchmark from Google data-set with cost of all solutions that are at (a) Hamming distance of one, or  (b) Hamming distance of two from the desired cuts.}
   \label{fig:1h2haway}
   \vspace{-0.05 in}
\end{figure}

\subsection{Hamming Clustering and QAOA}
We observe Hamming structure in the output of \texttt{QAOA} circuits, wherein most incorrect outcomes are close to the desired outputs in the Hamming space. However, despite this closeness in Hamming space, incorrect outcomes can significantly change the expected value compared to the noise-free expected value. To illustrate how just a few bit-flips can significantly degrade the average solution quality, we show a partial cost landscape of max-cut problem used for \texttt{QAOA-10} from Google dataset in Figure~\ref{fig:1h2haway} and overlay two desired cuts with the lowest cost. Note that the maxcut problem is formulated such that cost of the desire cut is negative~\cite{harrigan2021quantum}.

The staircase plot in Figure~\ref{fig:1h2haway}(a) shows the cost of all the solution strings that are one Hamming distance away from desired cuts, whereas Figure~\ref{fig:1h2haway}(b) represents the cost of all the solution strings that are two Hamming distance away from the desired cuts. Figure~\ref{fig:1h2haway}(a) show that the solutions that are just one Hamming distance away have 2x higher cost, and strings that are two Hamming distance away can have up to 10x higher cost compared to the desired solution as shown in the Figure~\ref{fig:1h2haway}(b).

\newpage
\section{Hamming Reconstruction}
HAMMER is a post-processing technique that leverages the observation about the erroneous outcomes being close in the Hamming space to the correct outcome to produce a modified distribution. In this section, we first provide an overview of HAMMER and then the details of each step.

\subsection{Overview of HAMMER}

The output of a NISQ program can be represented as a probability distribution, where the outcome $x_i$ occurs with probability $Pr(x_i)$ measured across all the trials. Unfortunately, correct outcomes are often indistinguishable from incorrect ones due to errors, and therefore, the probabilities associated with them are generally inaccurate and insufficient to infer the solution for large programs. For example, Figure~\ref{fig:hammerblockdiagram}(a) shows the output distribution ($\mathbf{P}_{in}$) of a 3-qubit circuit whose correct output is "\textit{111}". However, "\textit{111}" does not appear with the highest frequency. If we rely on the probabilities and pick the most frequent outcome as the program solution, we will incorrectly pick "\textit{000}". The goal of HAMMER is to \textit{accurately} estimate the likelihood $\mathbf{L}(x_i)$ of every outcome $x_i$ in $\mathbf{P}_{in}$. 

Figure~\ref{fig:hammerblockdiagram} provides an overview of HAMMER. HAMMER consists of three steps. The first step identifies the Hamming distance neighborhood for each unique outcome in the histogram. This is used to compute a "Neighbourhood Score". The second step analyzes the neighborhood scores of all unique outcomes to develop a "weight" that must be assigned to each neighborhood that is at a distance $K$. The third step is to use the weight and the probability distribution of the neighborhood to compute the effective value for each outcome in the probability distribution (and normalize).

Thus, HAMMER determines the \textit{likelihood} $\mathbf{L}(x_i)$ of every outcome $x_i$ being an error-free answer, by combining (a) its probability of occurrence and (b) a "Neighbourhood Score", $\mathbf{S}(x_i)$, as described in Equation~\eqref{eq:likelihood}. We design a robust Likelihood function that obtains the neighborhood score by exploiting the structure in the Hamming space. Using this score, the singleton outcomes that appear without structure in Hamming space are penalized, whereas outcomes with Hamming structure get boosted. We discuss each step of HAMMER next. 

\begin{equation}
    \label{eq:likelihood}
    \mathbf{L}(x_i) = Pr(x_i) \times \mathbf{S}(x_i)
\end{equation}

\vspace{0.01 in}

\subsection{Step-1: Create Hamming Spectrum}

\begin{figure}[htb]
\centering
    \includegraphics[width=0.99\columnwidth]{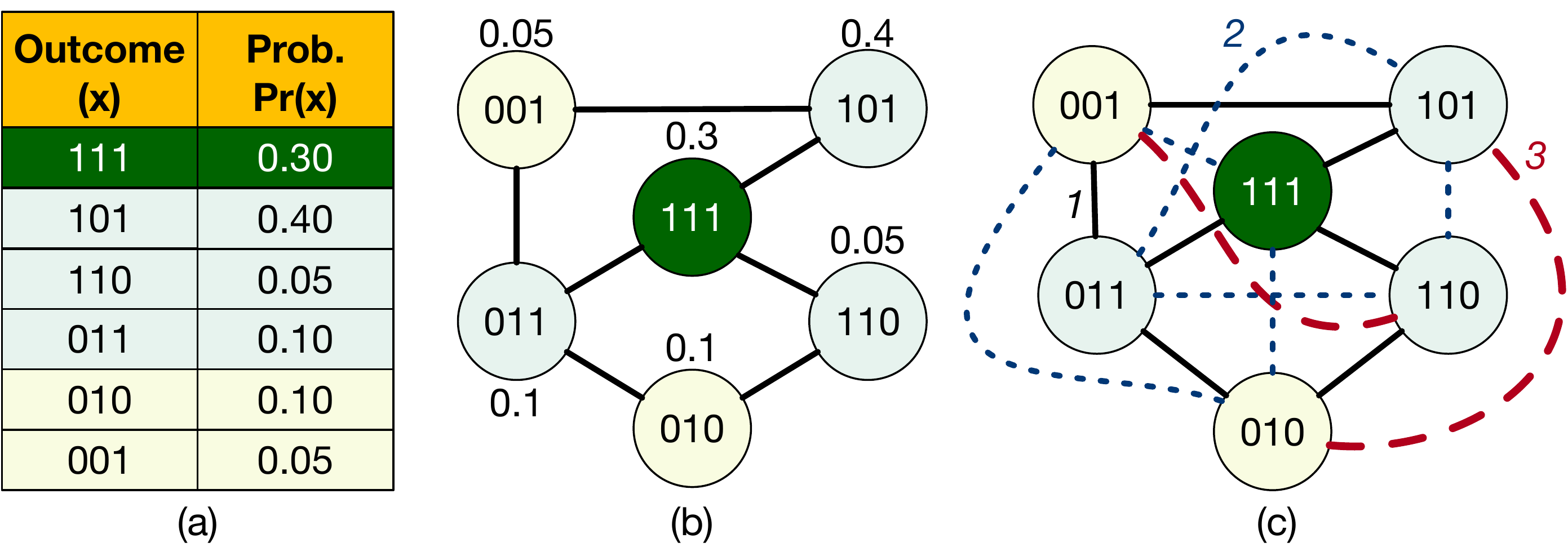}
    \vspace{0.0 in}
    \caption{(a) Output Probability Distribution. (b-c) Hamming Graph Representation of the Output Distribution.}
    \label{fig:pdftograph}
\end{figure}

To capture the Hamming structure, we introduced the notion of {\em Hamming Spectrum} in Section~\ref{sec:char}. We can also represent this structure by using an equivalent Hamming graph where individual outcomes are the nodes of the graph, and the weight of the edges connecting the nodes is the Hamming distance between them. For example, the output distribution in Figure~\ref{fig:pdftograph}(a) can be represented as a six node Hamming graph, as shown in Figure~\ref{fig:pdftograph}(b-c).

For simplicity of illustration, Figure~\ref{fig:pdftograph}(b) only shows those edges that are one Hamming distance away from each other, whereas, in reality, the graph has additional edges corresponding to other Hamming distances, as shown in Figure~\ref{fig:pdftograph}(c). In this example, although the correct outcome "111" occurs with a lower probability, by looking at the Hamming graph, we observe that "111" has more neighbors than the most frequent outcome "101".

To leverage this structure in deriving the likelihood function at scale, there are two key challenges:

\vspace{0.05 in}
\begin{enumerate}
    \item What is the right size of the neighborhood in Hamming space that will enable a robust likelihood function?

    \item How to compute the neighborhood score to boost program fidelity?
    
\end{enumerate}
\vspace{0.05 in}

\begin{figure*}[t]
\centering
    \includegraphics[width=1.99\columnwidth]{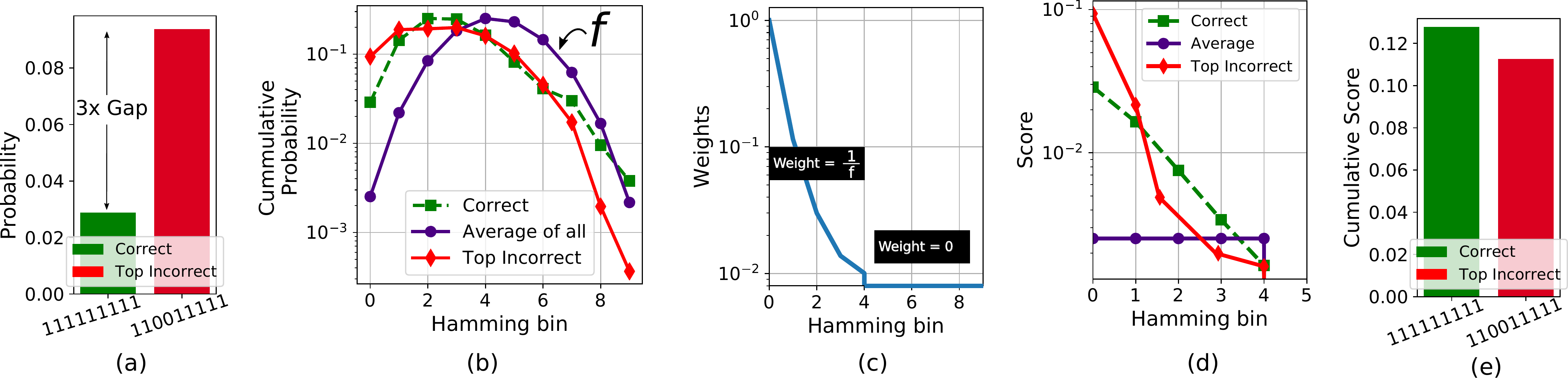}
\vspace{0.05 in}
    \caption{The probability distribution (P) produced by a \texttt{BV-10} circuit.  (a) Probability of correct and top incorrect outcomes in P. (b) Cumulative Hamming Spectrum (CHS). (c) Weights computed for P. (d) Neighbourhood Score for the correct, top incorrect, and average of all strings. (e) Cumulative Score.}
    \label{fig:designexample}
\end{figure*}

\vspace{0.1 in}
\textbf{Size of Neighborhood:} Our characterization data show that although the most frequent incorrect outcomes are close to the correct answers in the Hamming space, they disperse in multiple Hamming bins with increasing circuit depth. We observe that certain multi-bit flips become dominant in the output distribution, which increases the Expected Hamming Distance (EHD) and reduces the density of clusters. For example, the most frequent incorrect outcome "\textit{110011111}" of a 10-qubit \texttt{BV} circuit is two Hamming distance away from the correct answer "\textit{111111111}", as shown in Figure~\ref{fig:designexample}(a). 

Thus, to compute a robust neighborhood score, we cannot rely only on neighbors at Hamming distance of one but also need to consider the neighbors at slightly larger distances.

\vspace{0.1in }
\textbf{Neighbourhood Score:} The influence of a neighbor is computed from its Hamming Spectrum ($HS$). Increasing the neighborhood size causes the number of neighbors, and therefore its influence, to grow rapidly, as shown in Figure~\ref{fig:pdftograph}(c). For an $n$ qubit program, there are $n\choose k$ possible neighbors that are within "$k$" Hamming distance away. In the limiting case, when the entire neighborhood is considered, the neighborhood score will be influenced by all the other outcomes, eventually yielding a uniform score across all outcomes. Thus, there exists a trade-off between the neighborhood size and its influence. While looking at very small neighborhoods may not be enough, particularly for large circuits as in the case of the \texttt{BV-10} circuit example above, bigger neighborhoods can dilute the influence of the individual neighbors. 

\subsection{Step-2: Compute Per-Distance Weights} \label{sect:weighingHammNeighbours}

Ideally, we want to look at larger neighborhoods and simultaneously ensure that the neighborhood score derived from them has a positive influence on the likelihood function. To solve this problem, we limit the influence of each neighbor by assigning a  weight ($W$) to each neighbor based on its Hamming distance ($d$) while computing the neighborhood score. Quite often, an erroneous outcome that occurs with a very low probability may exist in a very influential neighborhood, such as outcome "001" in Figure~\ref{fig:pdftograph}.
To avoid such low probability outcomes deriving from a rich neighborhood, we introduce a filter function $\pi(x)$ that differentiates between dominant and average incorrect outcomes, as shown in Equation~\ref{eq:nscore} 
\begin{equation}
    \label{eq:nscore}
    \mathbf{S}(x) =\sum_{i=0}^{d} \pi(x)\times W_i \times CHS_i(x)
\end{equation}

Next, we discuss how to design the weights and the filter function to exploit the structure of errors in Hamming space. 

The outcomes generated on a NISQ machine may be grouped into (1) the correct outcome(s), (2) a few dominant incorrect outcome(s), and (3) a large number of average incorrect outcomes that appear with low frequency.
To distinguish between the three, we utilize the observation that dominant incorrect outcomes lie in close proximity to the correct answers in the Hamming space. To quantify this observation, we define the Cumulative Hamming Strength (CHS), a vector that holds the total probability of all the outcomes that are "d" Hamming distance away from a given outcome. For an $n$-qubit program with $n$-bit output, the possible Hamming distances range from 0 to $n$. For a given string, we compute the CHS by adding the probabilities of the outcomes in each individual Hamming bin of the HS. For example, Figure~\ref{fig:designexample}(b) shows the CHS of a \texttt{BV-10} program for the correct output, the most frequent incorrect outcome, and the average of all the outcomes. We observe that the CHS of the correct and dominant incorrect outcomes peaks at a low Hamming bin, as shown in  the~\ref{fig:designexample}(b). On the contrary, the average (incorrect) outcome peaks at $d=\frac{n}{2}$  in the Hamming spectrum. This observation is consistent in all our experiments and closely matches the theoretical estimate from a uniform error model as the total number of entries in the Hamming bin approximately scale as  $n\choose d$  whose maxima is at d= $\frac{n}{2}$. This suggests - (1) infrequent (average) incorrect outcomes exhibit weak structure in Hamming Space, (2) structure exists only for the dominant incorrect outcomes, and they appear in the close neighborhood of the correct answer. HAMMER uses this insight to differentiate between incorrect and correct outcomes.  

Since infrequent average outcomes constitute the majority of the output distribution, the CHS of the average case represents the global neighborhood information. As most outcomes are erroneous, we use the average CHS to compute the weights for estimating the neighborhood score. We compute the weights (W) by inverting the average CHS as shown in Figure~\ref{fig:designexample}(c). Furthermore, to prevent the infrequent outcomes benefiting from a rich neighborhood, we limit the neighborhood sizes up to $\frac{n}{2}$ by assigning zero weight for Hamming bins greater than $\frac{n}{2}$. Each outcome in the output distribution is assigned its individual Neighborhood Score $S(x)$ by multiplying its CHS and the Weights. Figure~\ref{fig:designexample}(d) shows the neighborhood score for the correct outcome, the dominant incorrect outcome, and the average outcome.

\vspace{-0.1 in}
\subsection{Step-3: Update the Probability Distribution}

HAMMER assigns the Neighborhood score by computing a dot product between the CHS vector and the weight vectors. However, in this process, a low probability string that is part of the rich neighborhoods gets assigned a high score. This equalization step can result in neighborhood scores that are very similar across all outcomes reducing the effectiveness of the HAMMER. For circuits with 10+ qubits, we observe that many low probability outputs are part of rich neighborhoods. For example, in the case of a \texttt{QAOA-16} benchmark, there are more than 1000 outcomes within the radius of three Hamming distance. With increasing circuit size, although the expected Hamming distance grows slowly, we see an explosion in the total number of possible outcomes. For an effective score update, it is essential to account for this effect, and we introduce a filter function to update the neighborhood score. 

The filter function determines if the string with low probability gets any credit from nodes that are high probability. For a given string (s), while computing neighborhood score, we only consider neighboring strings ($s_{i}$), which have a lower probability than the given string. This modulates the benefits given to the strings that are in the rich neighborhood but sampled with a high probability in the original distribution.

\subsection{Tying It All Together}
We use an example to show how HAMMER improves fidelity using neighborhood scores. Figure~\ref{fig:designexample}(a) shows the probability of two strings in the \texttt{BV-10} output distribution. The output string "111111111" is an error-free output, whereas "110011111" is an incorrect string that appears with the highest probability. HAMMER's goal is to boost program fidelity by reducing the gap between correct and most frequently occurring incorrect outcomes. 

In the baseline, there is a 3x gap between the two. Although the correct string has a low probability, it is part of a stronger neighborhood, as shown in ~\ref{fig:designexample}(b). In HAMMER, we compute the average CHS and invert it to generate the weights, as shown in Figure~\ref{fig:designexample}(c). We compute the neighborhood score for the correct output and the incorrect output by multiplying the CHS with the weights. Figure~\ref{fig:designexample}(d) show neighbourhood score for each bin. We sum the neighborhood scores to compute the total score for both the output strings, as shown in Figure~\ref{fig:designexample}(e). Using these post-processing steps, we generate the output probability distribution. With HAMMER, the correct string now appears with a higher probability compared to the strongest incorrect string. Algorithm \ref{alg:HAMMER} in the Appendix describes the overall steps for HAMMER.

\section{Evaluation Methodology}
We evaluate the effectiveness of HAMMER by running various benchmarks on three IBM quantum computers. Furthermore, we test HAMMER on publicly available quantum datasets from Google~\cite{Googlefigshare}. 

\subsection{Benchmarks and Datasets}

{\noindent \textbf{Google Dataset:}} We test HAMMER on the Google dataset (see Table~\ref{tab:gbenchmarks}). The output distributions in this dataset are collected by running {\em Quantum Approximate Optimization Algorithm (\texttt{QAOA})} instances on the 53-qubit Sycamore processor~\cite{sycamoredatasheet}. These \texttt{QAOA} circuits are used to find the max-cut on Grid, Sherrington-Kirkpatrick and 3-regular input graphs~\cite{Googlefigshare}.

\begin{table}[htb]
\begin{center}
\begin{small}

\caption{Benchmarks from Google Dataset~\cite{Googlefigshare}}
\vspace{-0.15 in}
\setlength{\tabcolsep}{1.0mm} 
\renewcommand{\arraystretch}{1.4}
\label{tab:gbenchmarks}
{\footnotesize
\begin{tabular}{ |c|c|c|c|c|c|} 
\hline
\multirow{2}{*}{Name} & \multirow{2}{*}{Algorithm details} & \#Qubits  & P & Total & Figure of \\
 & &  (n) & Layers & Circuits & Merit \\
\hline
\texttt{QAOA}  & Maxcut on Grid  & 6–20 & 1 to 5 & 120 & CR \\

\hline

\texttt{QAOA} &  Maxcut  on 3-Reg Graphs & 4–16 & 1 to 3 & 200 & CR \\

\hline
\end{tabular}}
\end{small}
\end{center}
\end{table}

{\noindent \textbf{IBM Dataset:}} We use quantum benchmarks of different sizes and structures to evaluate our proposed design. Table~\ref{tab:ibenchmarks} summarizes the benchmarks used in this paper. For the \texttt{QAOA}  benchmarks, we use it in the context of Max-Cut problems on a wide range of random and regular graphs. The graphs are generated using the Erdos-Renyi method~\cite{bollobas2001random}. To obtain a wide range of graphs, we vary the degree of connectivity between 0.2 (sparse) to 0.8 (highly connected), depending on the size of the problem. We adopt this approach from prior works focused on \texttt{QAOA} circuits~\cite{alam2020accelerating,gokhale2019partial}.

\begin{table}[htb]
\begin{center}
\begin{small}

\caption{Details of NISQ benchmarks on IBM Machines}
\vspace{-0.15 in}
\setlength{\tabcolsep}{1.0mm} 
\renewcommand{\arraystretch}{1.4}
\label{tab:ibenchmarks}
{\footnotesize
\begin{tabular}{ |c|c|c|c|c|c|} 
\hline
\multirow{2}{*}{Name} & \multirow{2}{*}{Algorithm details} & \#Qubits  & P & Total & Figure of \\
 & &  (n) & Layers & Circuits &Merit \\
\hline 
\hline
\texttt{BV}  & Bernstein-Vazirani & 5–15 & - & 88 & IST, PST \\

\hline
\texttt{QAOA} &  Maxcut  on 3-Reg Graphs  & 5–20 & 2 and 4  & 70 & CR, PF \\

\hline
\texttt{QAOA}  &  Maxcut Rand Graphs   & 5–20 & 2 and 4 & 70 & CR, PF \\

\hline
\end{tabular}}
\end{small}
\end{center}
\end{table}

\subsection{Compiler and Hardware}
We use the Qiskit compiler tool-chain from IBM~\cite{qiskit,li2018tackling,murali2019noise}. Additionally, we perform the compilation step recursively to ensure minimum number of \texttt{CNOT}s. For all evaluations, we use three real quantum hardware from IBM. Note that although all of these machines have a Quantum Volume~\cite{quantumvolume} of 32 they have very different error characteristics. Generally, NISQ programs are executed multiple times (called trials) and by default, IBMQ systems use 8K trials. For our evaluations, we execute between 8K-32K trials. This serves as our \textit{\textbf{baseline}} for evaluation. We also evaluate HAMMER across multiple calibration cycles and observe similar results.

\newpage
\section{Evaluations}

We now discuss the effectiveness of HAMMER in improving the quality of solution for key NISQ benchmark circuits. 

\subsection{Figure of Merit for BV}
We use two different metrics derived from prior works to evaluate our design, and these metrics are discussed below:

\vspace{0.05in}
\noindent \textit{\textbf{(1)  Probability of Successful Trial (PST):}} measures the probability of the correct answer and is defined as the ratio of the number of trials that produce the correct outcome(s) to the total number of trials, as described in Equation~\eqref{eq:pst}. We use PST for \texttt{BV} circuits as it has one correct solution. 

\begin{equation}\label{eq:pst}
PST = \frac{\textrm{Number of Correct Trials}}{\textrm{Total Number of Trials}}
\end{equation}

\vspace{0.05in}
\noindent \textit{\textbf{(2) Inference Strength (IST)}} is used to account for the magnitude of both the correct and the incorrect answers. IST is the ratio of the frequency of the correct output to the frequency of the most commonly occurring erroneous output, as described in Equation~\eqref{eq:ist}. If IST exceeds 1, the system will be able to correctly infer the output, whereas if IST is significantly lower than 1, then the wrong answer(s) would mask out the correct answer. 

\begin{equation}\label{eq:ist}
IST = \frac{Pr(S_{correct})}{Pr(S_{incorrect})}
\end{equation}

\begin{figure}[htb]
\centering
    \includegraphics[width=1.0\columnwidth]{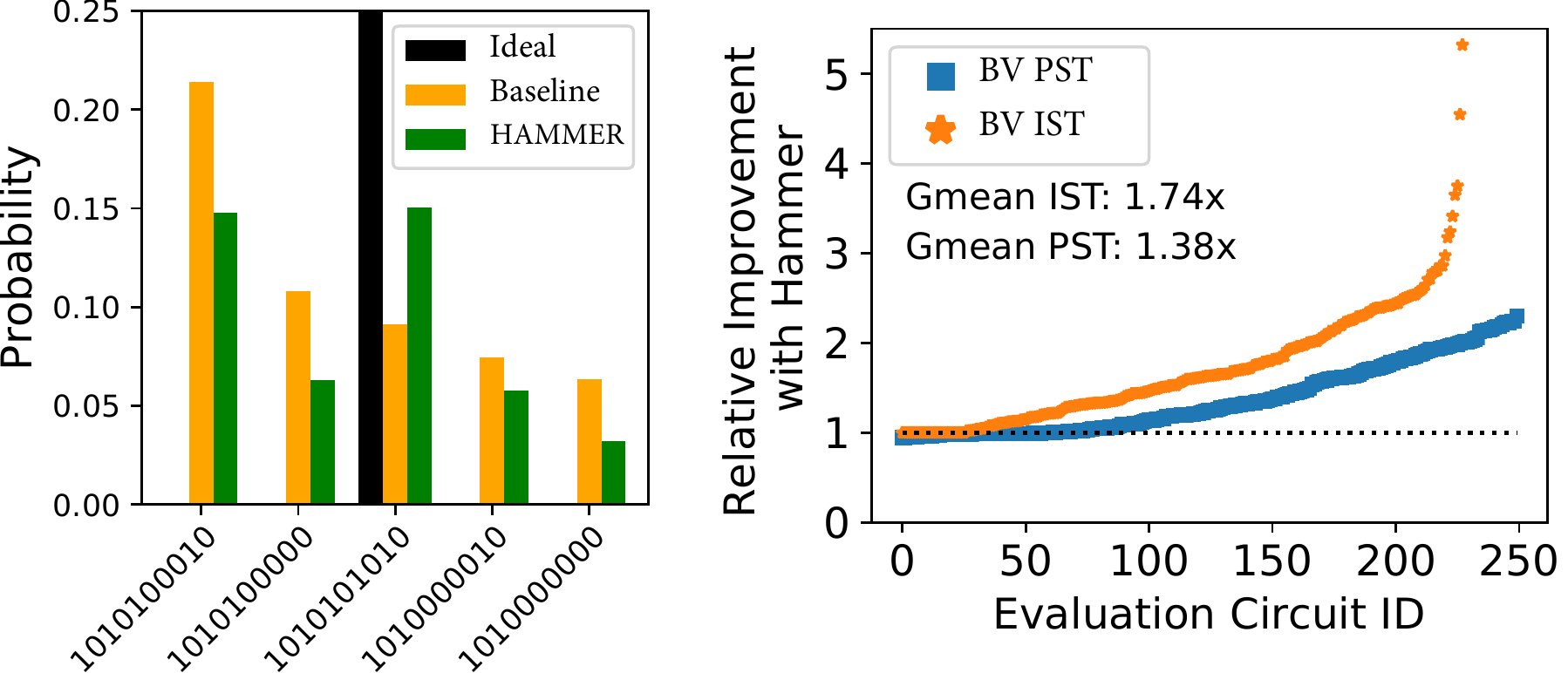}
    \caption{(a) Output of BV-10 circuit with \textit{1010101010} key  (b) Improvement in PST and IST with HAMMER for 250 BV circuits with 5-16 qubits over baseline on three IBM Machines.}
    \label{fig:bv_results}
\end{figure}

\subsection{Impact on Bernstein Vazirani Circuits}

Figure~\ref{fig:bv_results}(a) shows an output of a \texttt{BV-10} circuit executed on a noiseless simulator, on IBM-Paris (baseline), and a post-processed output from HAMMER. For noiseless simulation, the probability of solution string "1010101010" is 100\%, whereas for the probability drops to 8\% in the baseline. Furthermore, the incorrect outcome "1010100010" is the most frequent string with 20\% probability in the baseline. The resultant distribution obtained by applying HAMMER on the output distribution from the baseline is shown in Figure~\ref{fig:bv_results}(a). HAMMER improves the PST from 8\% to 14\% by boosting the correct solution string as it is more likely to be a correct answer based on the Hamming structure. Moreover, HAMMER attenuates the dominant incorrect outcome and increases the IST from 0.4 to 1.01, such that the solution key has the highest frequency of occurrence. We observe a similar boost in program fidelity with HAMMER for 250 BV circuits. Figure~\ref{fig:bv_results}(b) shows the relative improvement in PST and IST. We observe on average, HAMMER improves fidelity by 1.38x and up to 2x. At the same time, it boosts IST by up to 5x and, on average, by 1.74x.

\begin{figure*}[t!]
\centering
    \includegraphics[width=2\columnwidth]{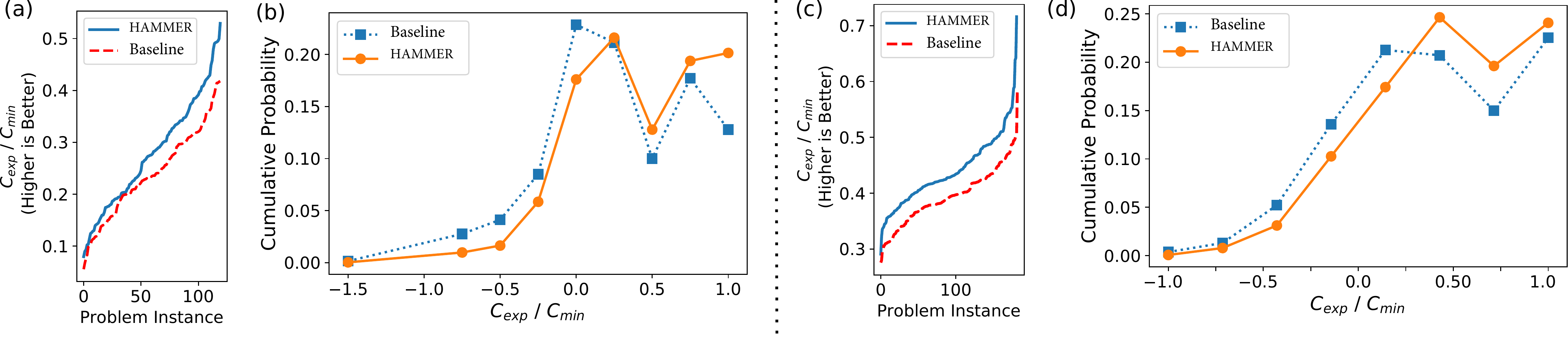}
    \vspace{-0.1in}
    \caption{(a) Cost Ratio S-Curve for Baseline and HAMMER  for 3-Reg input graphs. (b) Output for \texttt{QAOA-10} example on 3-Reg input. (c) Cost Ratio S-Curve for Baseline and HAMMER for grid input graphs. (d) Output for \texttt{QAOA-12} example on grid input.}
    \label{fig:qaoa_results}
    \vspace{-0.1in}
\end{figure*}

\subsection{Figure of Merit for QAOA Circuits}

\noindent \textit{\textbf{Cost Ratio (CR):}} 
Hybrid quantum-classical algorithms such as QAOA use parametric circuits with a classical optimization loop, in which the circuit parameters are tuned to minimize the cost function. To calculate the cost, the circuit is first executed on the NISQ hardware for thousands of trials, which generates an output distribution that consists of possible sample solution strings and their probabilities. We can evaluate the expectation value of the cost function by computing the weighted average of cost corresponding to each solution string in the output distribution~\cite{harrigan2021quantum,qaoa}.

\begin{equation}
CR = \frac{C_{exp}}{C_{min}}
\label{eq:arg}
\end{equation}

We analyze the impact of errors on the effectiveness of QAOA while evaluating the expectation value and discuss the efficiency of HAMMER in tackling these errors. We compute the \textbf{\textit{ Cost Ratio (CR)}} which is the ratio of the average quality ({$C_{exp}$}) and the lowest possible cost of the solution ({$C_{min}$}), as described in Equation~\eqref{eq:arg}. A higher CR indicates the better average quality of the solution. In the classical component of these algorithms, an optimizer uses the expectation values to tune the circuit parameters and eventually converges on a distribution that maximizes the expectation value and allows us to estimate the high-quality solution of the problem. Unfortunately, hardware errors that result in noisy distributions and low-quality expected values can disrupt the training process~\cite{arute2020quantum}.

\subsection{Impact on Quality of QAOA Solutions}

\textbf{Results on Google Dataset:} We run HAMMER on 320 QAOA circuits used to solve Max-Cut problems with 200 \textit{Grid} and 120 \textit{3-Regular} input graphs to quantify improvements in CR from Google dataset~\cite{Googlefigshare}. The baseline data uses a post-measurement correction scheme to reduce the readout bias~\cite{harrigan2021quantum}. Figure~\ref{fig:qaoa_results}(a) shows  an S-curve for the Cost Ratio (CR) of the baseline and HAMMER data for 3-Regular graphs from 6 to 16 nodes and for $p$ = 1 to 3 layers. Ideally, without noise, the CR ranges from 0.7 to 0.9, only depending on the number of layers. However, on Google Sycamore, qubit errors reduce the average quality of solution $C_{exp}$ and CR drops to 0.08 to 0.4, as shown in Figure~\ref{fig:qaoa_results}(a). HAMMER boosts the CR consistently for all input circuits showing improvements up to 2.4x in the CR. Figure~\ref{fig:qaoa_results}(b) show an output distribution of a QAOA-10 circuit with p=2 layers. We plot the cumulative probability of all solutions (y-axis) corresponding to a Ratio of ${C_{sol}}/{C_{min}}$ value on x-axis. The quality of the solution is highest when ${C_{sol}}={C_{min}}$ , and it degrades with decreasing value of the ratio. Note that  $\frac{C_{sol}}{C_{min}}$ can be negative, which represents a sub-optimal cut. We want to maximize the probability of all solution strings with higher $\frac{C_{sol}}{C_{min}}$ value to boost the average cost $C_{exp}$. In Figure~\ref{fig:qaoa_results}(b), HAMMER achieves this goal by increasing the cumulative probability from 12\% to 19.5\% for optimal cuts and reducing the probability of sub-optimal cuts. We observe a similar trend for grid input graphs in  Figure~\ref{fig:qaoa_results}(c) and (d). For Grid graph inputs, both baseline and HAMMER have higher CR due to reduced circuit depth and total gate counts for grid circuits that do not require SWAPs.    

\vspace{0.05in}

{\noindent \textbf{Results on IBM Dataset:}} We use HAMMER to post-process 140 QAOA circuits executed on three IBMQ systems. For these benchmarks, we observe total variational distance (TVD) decreases by 1.23x and CR increases by 1.39x on average. We evaluate TVD by comparing output obtained on the hardware and ideal simulation of the circuit.

\subsection{Reclaiming Algorithmic Benefits of QAOA}

\begin{figure*}[t]
\centering
    \includegraphics[width=2\columnwidth]{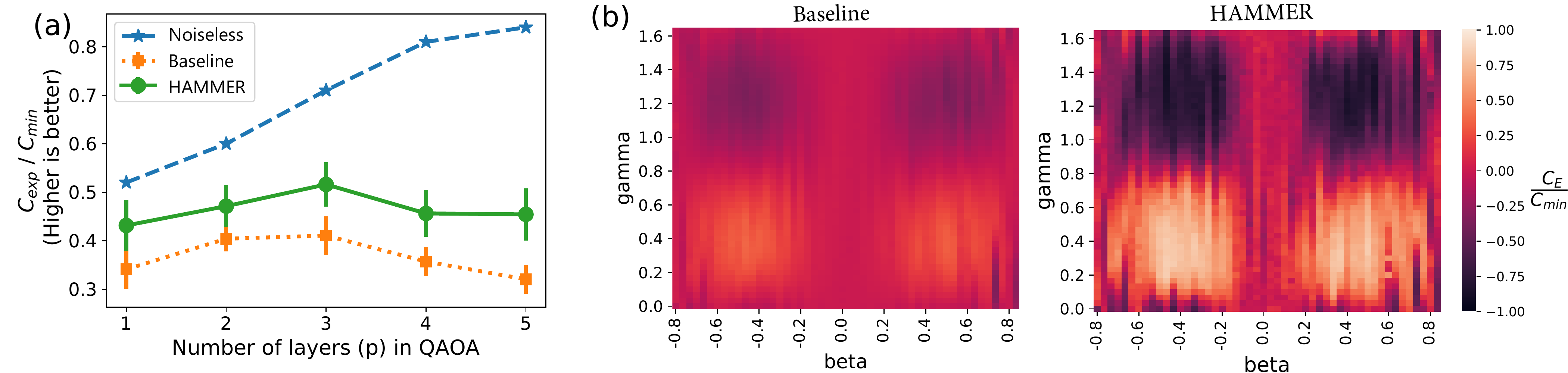}
    \caption{(a) Quality of solution with increasing layers in QAOA circuits solving Max-Cut for grid graphs with 6-20 nodes, with total 200 circuits. (b) Optimization landscape for 3-Regular graph for Google baseline and HAMMER. }
    \label{fig:qaoa_land}
\end{figure*}

In theory, the solution quality of QAOA improves rapidly with the increasing number of "p" layers. However, by increasing the number of layers, QAOA circuits executed on NISQ hardware become more error-prone and produce noisy sub-optimal solutions with a higher frequency. This degrades the average quality of the solution. Furthermore, training a QAOA circuit with a higher "p" becomes challenging due to increasing noise. Figure~\ref{fig:qaoa_land} shows CR for Google baseline data with an increasing value of "p" for finding max-cut on the grid of 10 to 20 nodes. In the ideal noiseless case, the quality of solution, i.e. CR, improves monotonically with increasing "p" number of layers. Whereas, on Google's Sycamore device, the quality of solution peaks at p=2 and reduces. When we post-process Google data with HAMMER, we observe a peak in quality at p=3. As HAMMER can suppress hardware noise, it can reclaim algorithm benefits of QAOA at higher "p" values.            

Moreover, to understand if HAMMER can improve the classical optimization steps in variational mode, we run HAMMER on a complete landscape of \texttt{QAOA-14}, solving maxcut for 3-regular graph. We observe that HAMMER consistently enhances the quality of solution for each data point on the grid and sharpens the gradient.

\subsection{Complexity Analysis of HAMMER}
For a given quantum program with $n$ qubits, let $N$ be the number of non-zero entries in the (noisy) distribution obtained on a NISQ computer. From the memory complexity viewpoint, HAMMER requires to store two vectors of size $n/2$  for storing Hamming score (HS) and weight vectors—denoted by ${HS}$ and ${W}$ in Algorithm \ref{alg:HAMMER}, respectively.
Hence, the memory required by HAMMER grows linearly with the number of qubits—i.e., $O({n})$. Our analysis shows that the memory required is less than 1 MB even for problems using up to 500 qubits. 
For computational complexity, HAMMER performs $N^2+N$ steps to compute the Hamming weight vector, $N^2$ steps for computing the likelihood of observations being a correct outcome, and ${N}$ steps for normalizing likelihoods. Thus, the execution time of HAMMER grows quadratic with the number of unique outcomes - $O(N^2).$ 
 For HAMMER, the execution time gets determined by the number of unique outcomes, which is limited by the number of trials for large programs. 

For example, Google uses a total of 25,000 trials irrespective of the size of the \texttt{QAOA} problem~\cite{harrigan2021quantum}. The largest \texttt{QAOA} instance that we evaluate with HAMMER had 24 qubits. For this instance, there were about 20K unique outcomes, and HAMMER required 56 seconds for a Python-based single-threaded code. On NISQ machines, we expect to observe a limited number of unique outcomes. Even if we use a few thousand trials and each trial produces a unique outcome, HAMMER can process this within a few minutes. Table~\ref{tab:scale} shows the complexity of HAMMER for machines with up-to 500 qubits when we run 32K or 256K trials. The time complexity of HAMMER is small, even with 256K unique outcomes. 

\ignore{
\begin{table}[htb]
\begin{center}
\begin{small}
\caption{Scalability Analysis of HAMMER }
\vspace{-0.15 in}
\label{tab:scale}
\label{tab:scalabilitynumbers}
{
\begin{tabular}{ |c|c|c|c|c|} 
\hline
{Qubits} & \multirow{2}{*}{Trials (T)} & Unique & \multicolumn{2}{c|}{HAMMER}  \\
\cline{4-5}
(n) & & Outcome  & Memory  & Operations   \\
\hline
\hline

\multirow{4}{*}{100} & \multirow{2}{*}{32K} &  10\% & $< 1MB$ & 0.001 Bln  \\
 \cline{3-5} 
& &  100\% & $< 1MB$& 1 Bln \\
 \cline{2-5} 
& \multirow{2}{*}{256 K} &  10\%  & $< 1MB$ & 0.6 Bln \\
 \cline{3-5} 
 & &  100\%  & $< 1MB$ & 64 Bln \\ \hline \hline

\multirow{4}{*}{500} & \multirow{2}{*}{32K} &  10\% & $< 1MB$ & 0.001 Bln  \\
 \cline{3-5} 
& &  100\% & $< 1MB$& 1 Bln \\
 \cline{2-5} 
& \multirow{2}{*}{256 K} &  10\%  & $< 1MB$ & 0.6 Bln \\
 \cline{3-5} 
 & &  100\%  & $< 1MB$ & 64 Bln \\ \hline

\end{tabular}}
\end{small}
\end{center}
\end{table}}

\begin{table}[htb]
\begin{center}
\begin{small}
\vspace{-0.1 in}
\caption{Number of Operations Required }
\vspace{-0.1 in}
\label{tab:scale}
\label{tab:scalabilitynumbers}
{
\begin{tabular}{ |c|c|c|c|} 
\hline
Trials & Unique & \multicolumn{2}{c|}{Operations in Billion}  \\
\cline{3-4}
(T) & Outcomes & Qubits (n) = 100  & Qubits (n) = 500  \\
\hline
\hline
\multirow{2}{*}{32K} & 10\% & 0.001 & 0.001 \\
\cline{2-4}
& 100\% & 1 & 1 \\ 
\hline
\multirow{2}{*}{256K} & 10\% & 0.6 & 0.6 \\
\cline{2-4}
& 100\% & 64 & 64 \\ 
\hline

\end{tabular}}
\end{small}
\end{center}
\vspace{-0.1in}
\end{table}

\section{Impact of Entanglement and Circuit Size on Hamming Behaviour}
\label{sec:entangle}
Quantum algorithms leverage entangled or correlated states to enable speedup over classical methods. Unfortunately, entangled states are prone to errors. Moreover, certain errors can rapidly spread among entangled qubits reducing the Hamming structure. To understand if the Hamming behavior persists with increasing entanglement, we run over a thousand benchmark circuits with varying degrees of entanglement on IBM quantum hardware. We use circuits with the following structure:  

\vspace{0.05 in}
\begin{center}
\begin{quantikz} \lstick{$\ket{0}^{\otimes n}$} & \gate{H} \qwbundle[ alternate]{} & \gate{U_{R}^{}} \qwbundle[ alternate]{} & \gate{U_{R}^\dag} \qwbundle[ alternate]{}&\gate{H} \qwbundle[ alternate]{} & \qwbundle[ alternate]{} \end{quantikz}
\end{center}
\vspace{0.05 in}

\begin{figure*}[t!]
\begin{subfigure}[b]{.245\textwidth}
\centering
\includegraphics[width=1.5in]{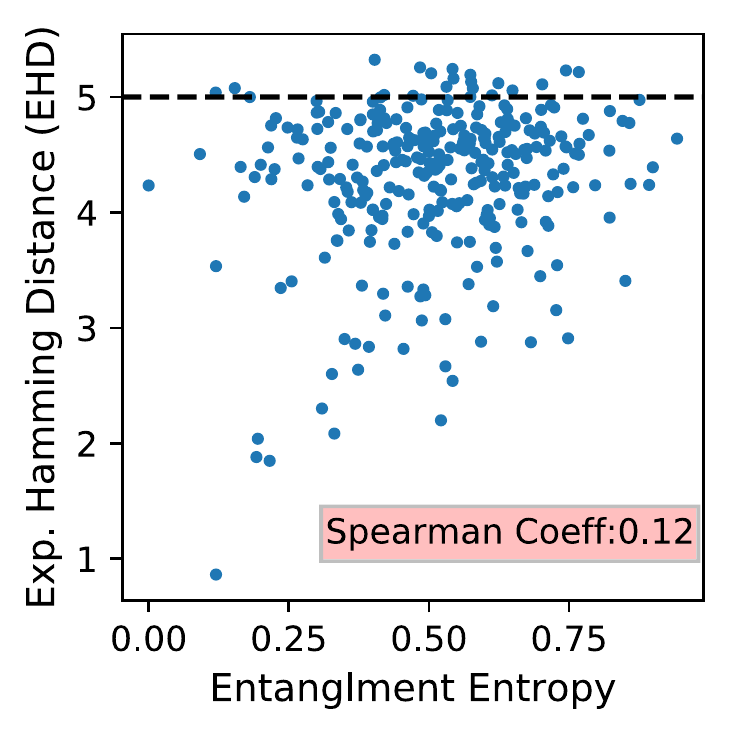}
\caption{}\label{fig:EHD1}
\end{subfigure}
\begin{subfigure}[b]{.245\textwidth}
\centering
\includegraphics[width=1.5in]{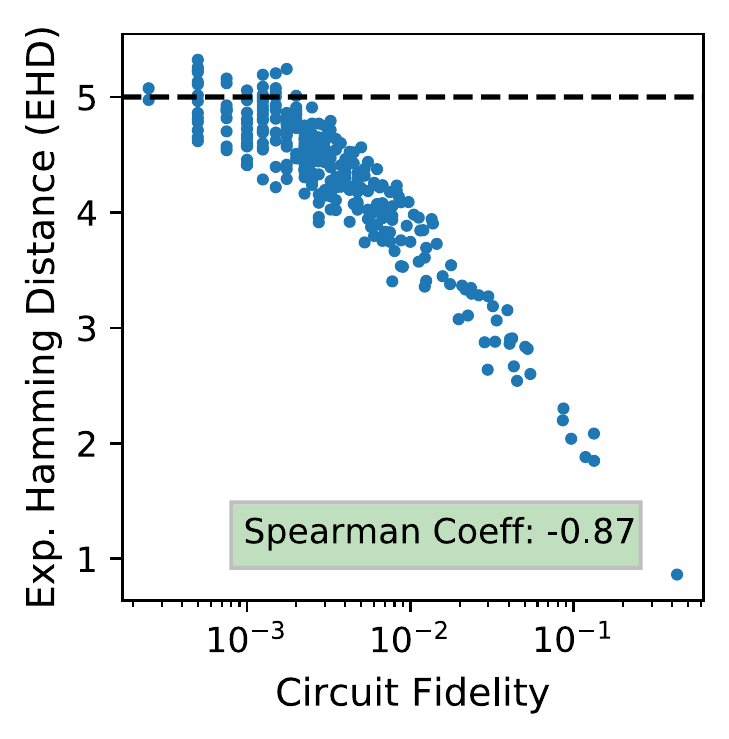}
\caption{}\label{fig:EHD2}
\end{subfigure}
\begin{subfigure}[b]{.245\textwidth}
\centering
\includegraphics[width=1.5in]{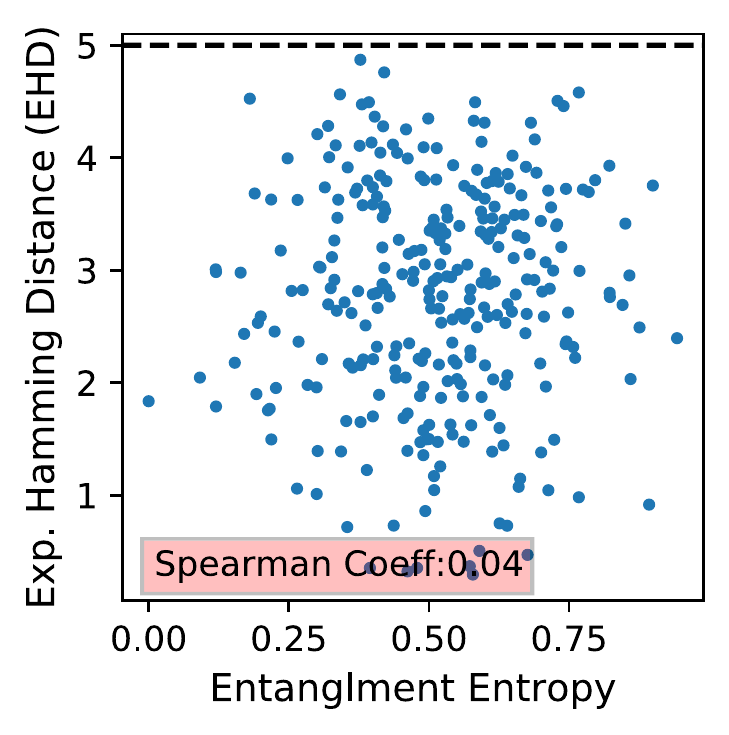}
\caption{}\label{fig:EHD3}
\end{subfigure}
\begin{subfigure}[b]{.245\textwidth}
\centering
\includegraphics[width=1.5in]{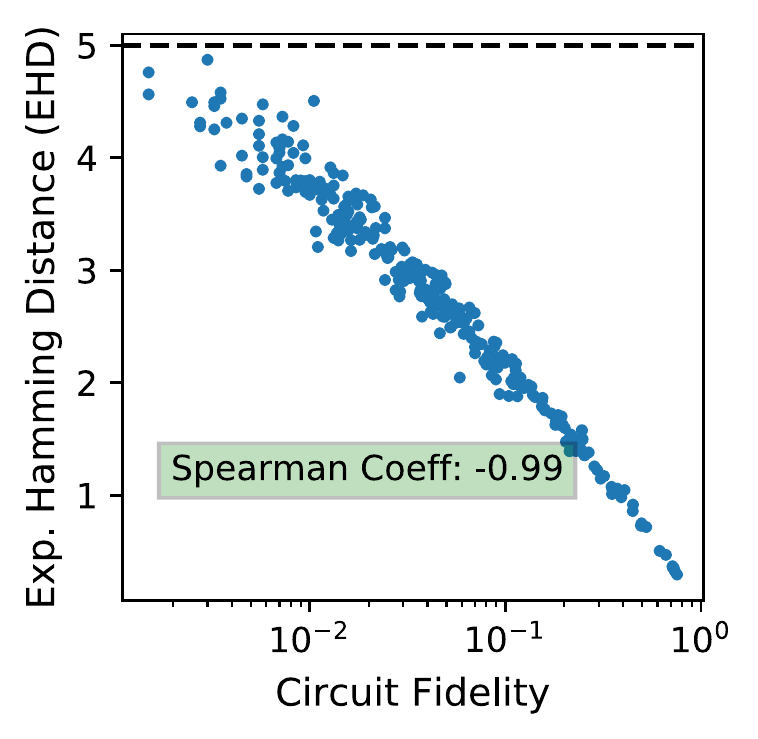}
\caption{}\label{fig:EHD4}
\end{subfigure}
 \caption{ EHD of high depth benchmark circuits with varying (a) Entanglement Entropy (b) Fidelity. EHD of low benchmark circuits with varying (c) Entanglement Entropy (d) Fidelity. Spearman Correlation Coefficient used to quantify the correlation.}
 \vspace{-0.1in}
\label{fig:EHD}

\end{figure*}

Each circuit starts and ends with a layer of Hadamard gates between which a random unitary,  $U_{R}$, and its reverse, $U_{R}^\dag$, are used. The sub-circuit $U_{R}$ comprises of randomly selected single-qubit (\texttt{Rz, Rx, Ry}) and two-qubit (\texttt{CX, CZ}) gates. The $U_{R}^\dag$ is the reverse circuit such that $U_{R}*U_{R}^\dag = I$. These circuits create an entangled state and gradually untangle it, producing an all-zero state ($\ket{0000..0_{n}}$). We use random unitaries to generate circuits with varying entanglement. Furthermore, the all-zero output state enables high fidelity measurements. We evaluate the degree of entanglement for a benchmark circuit by computing the entanglement entropy of the state produced by the sub-circuit : $H.U_{R}$ using ideal simulations. We run two sets of benchmark circuits - (1) high depth circuits with depth up to 25  (2) low depth circuits with depth up to 15.

Figure~\ref{fig:EHD}(a) shows how EHD changes with varying entanglement entropy for 300 ten-qubit circuits running on IBM hardware. We observe a weak correlation between entanglement entropy and EHD (Spearman coefficient: 0.2). Furthermore, EHD is lower than the uniform error model (dotted line), showing a strong Hamming structure. We observe a similar trend across different benchmark circuits. In fact, for shallower circuits shown in Figure~\ref{fig:EHD}(c), the correlation between entanglement and structure in errors weakens.

\begin{figure}[htb]
\centering
    \vspace{-0.1 in}
    \includegraphics[width=0.9\columnwidth]{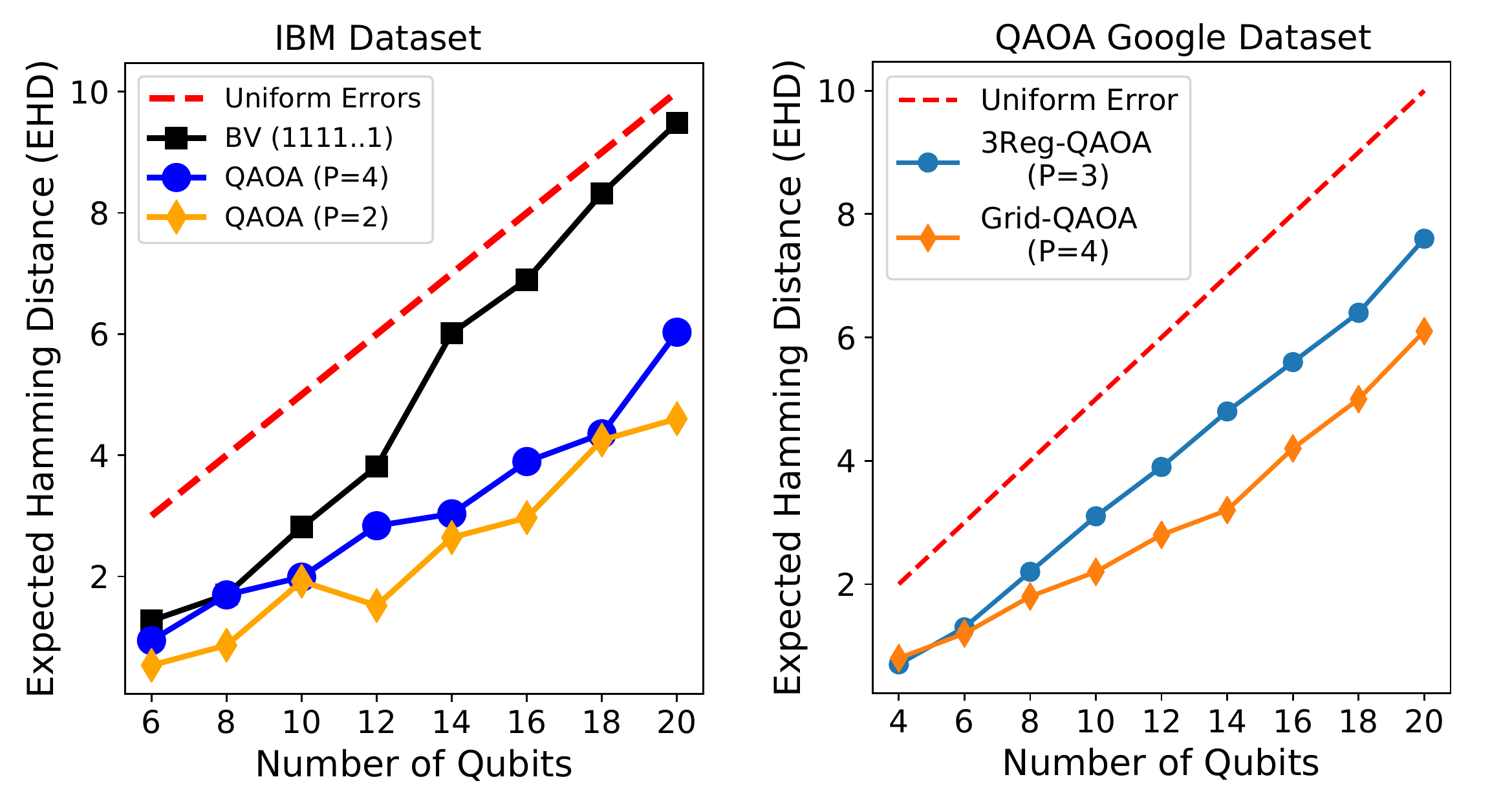} 
    \vspace{-0.1 in}
    \caption{EHD of output (a) IBM-Paris (b) Google-Sycamore}
    \label{fig:ehd_trend}
   \vspace{-0.05 in}
\end{figure}

While Hamming structure persists despite the increasing entanglement entropy or degree of entanglement, it reduces with increasing noise. As shown in Figure~\ref{fig:EHD}(b) and Figure~\ref{fig:EHD}(d) with decreasing fidelity, EHD increases for both low and high depth benchmark circuits. Increasing the size of quantum circuits increases the total number of operations and the duration for which qubits are active. This exacerbates the error rate, and with increasing errors, fidelity drops, and EHD increases. Similarly, Figure~\ref{fig:ehd_trend}(a) shows EHD for \texttt{BV} and \texttt{QAOA} circuits with 6 to 20 qubits. With the increasing size of circuits, we see an increase in EHD.  

With more errors, more incorrect answers are produced and scattered across the Hamming space. This results in a higher average Hamming distance between any two outcomes. However, compared to the uniform distribution where EHD is $\frac{n}{2}$, circuits discussed in  Figure~\ref{fig:ehd_trend} have significantly lower EHD. Furthermore, we observe different rates of increase in EHD for different circuits. BV circuits, for example, lose the structure much faster as compared to the QAOA circuits. This is because the depth of BV circuits increases super linearly compared to the linear increase in QAOA circuits.

Figure~\ref{fig:ehd_trend} highlights that circuit depth is a dominant factor that influences the EHD and the Hamming structure. Moreover, we observe this trend for over 400 QAOA circuits that are used to solve a max-cut problem on 3-regular, 2-regular, and Erdos-Renyi graphs with 6 to 20 nodes, as shown in Figure~\ref{fig:ehd_trend}(a), and a similar trend manifests for QAOA on Google-Sycamore, as shown in  Figure~\ref{fig:ehd_trend}(b). We observe that majority of the QAOA circuits have lower EHD than the uniform distribution, and for QAOA circuits with reasonable program fidelity, we observe a low EHD and dense clusters around the correct outcome. Our observations are consistent across three IBM machines and over twenty days of experiments. We also verify if the trend in error structure changes significantly due to qubit mapping. Our results suggest that although average fidelity and dominant incorrect answers change, the clustering effects remain similar across different mappings.

While we cannot establish the exact reasons for the Hamming behavior of incorrect outcomes, we can surmise a few reasons for why this behavior occurs in practice. Our evaluations show for shallow circuits, errors can have a localized impact preserving Hamming structure. Thus, shallower circuits can be expected to provide more Hamming structure than deeper circuits. In fact, given the constraints of NISQ systems, we are likely to run programs with shallow depth (for example, on IBM hardware, the fidelity of QAOA circuits drops below 1\% beyond circuit depth of 40). Our experiments show that for wide and low depth circuits, the errors are clustered in Hamming space.

\section{Related Work}

NISQ computers promise computational advantages for practical applications~\cite{qaoa,vqe,orus2019quantum}. In the absence of quantum error correction on these systems, software policies will play a vital role in closing the gap between the devices and the algorithms~\cite{chong2018closing,NAE}. Therefore, mitigation of hardware errors through software techniques is an active area of research. We can broadly classify them as compiler and post-processing techniques.

\vspace{0.05 in}
{\noindent \textbf{Compiler-Based Error-Mitigation:} }These methods focus on (1) generating highly optimized program schedules by accounting for the application-specific characteristics to reduce the circuit depth and number of operations~\cite{gokhale2019partial,gokhale2020optimized,li2018tackling,shi2019optimized,zulehner2018efficient,alam2020circuit} and (2) hardware error characteristics to perform noise-aware computations~\cite{noiseadaptive,murali2020software,micro1,FNM,tannu2019not,patel2020ureqa,patel2021qraft}. There are other approaches that decompose a circuit into smaller circuits and obtain the output distribution using a tensor product~\cite{tang2021cutqc}. Generally speaking, all of these policies focus on reducing the likelihood of a program encountering errors. Some of them particularly focus on a very specific source of error such as crosstalk between ongoing CNOT operations~\cite{murali2020software}, measurement errors~\cite{matrixmeasurementmitigation,bravyi2020mitigating,kwon2020hybrid,FNM,barron2020measurement,patel2020disq}, correlated errors~\cite{micro1}, or idling errors~\cite{smith2021error}. HAMMER uses a completely different approach and is compatible with all of these policies as it can be applied to a program compiled with these optimizations.  

\vspace{0.05 in}
{\noindent\textbf{Post-processing for Error-Mitigation:}} Recent work~\cite{patel2020veritas} has looked at the problem of correlated errors and proposed diverse mappings on different machines to reduce the magnitude of correlated errors. These schemes post-process the noisy outputs obtained from the individual mappings to offer a more accurate output distribution. Post-processing schemes have also been effective in mitigating measurement errors. These schemes use characterization data to compute a noise matrix which is then used to post-process the noisy output distribution obtained on a NISQ device~\cite{matrixmeasurementmitigation,bravyi2020mitigating}.  

\section{Conclusion}

We propose {\em Hamming Reconstruction (HAMMER)}, a post processing technique to boost the fidelity of a quantum program.  It uses the insight that correct outcomes are clustered in the Hamming space. HAMMER estimates the likelihood of each outcome based on a neighborhood of answers within a small Hamming distance of given answer. We evaluate the effectiveness of HAMMER using more than 500 key quantum benchmarks on IBM and Google Sycamore Datasets and show that HAMMER improves the quality of solution by 1.37x on average.  We also evaluate the scalability of HAMMER and show that it easily scales to machines with thousands of qubits.







\begin{acks}
We thank Matthew Harrigan from the Google Quantum AI Group for providing us the QAOA datasets used for some of the  evaluations in this paper. 
Poulami Das was funded by the Microsoft Research PhD Fellowship. Ramin Ayanzadeh was supported by the NSF Computing Innovation Fellows (CI-Fellows) program. 
This research used resources of the Oak Ridge Leadership Computing Facility at the Oak Ridge National Laboratory, which is supported by the Office of Science of the U.S. Department of Energy under Contract No. DE-AC05-00OR22725.
\end{acks}

\section*{Appendix A: Hamming Reconstruction}
\label{sec:bayesianalgorithm}
The Hamming Reconstruction algorithm is described below. 
\setlength{\textfloatsep}{0pt}
\SetAlgoNoLine
{\small
\begin{algorithm}[htb]
\caption{Hamming Reconstruction}
\label{alg:HAMMER}
\SetKwInput{KwInput}{Input}                
\SetKwInput{KwOutput}{Output}              
\DontPrintSemicolon
  
    \KwInput{ $P_{\text{in}}$ (\textrm{input distribution}), $n$ (\textrm{number of qubits}) }\vspace{0.0 in}
    \KwOutput{ $P_{\text{out}}$ (\textrm{output distribution from HAMMER})}\vspace{0.05 in}

  \SetKwFunction{CHE}{Create\_Hamming\_Envelope}
  \SetKwFunction{CHS}{Create\_Hamming\_Spectrum}
  \SetKwFunction{CWT}{Create\_Weight\_Table}
  \SetKwFunction{CBT}{Create\_Bonus\_Table}
  \SetKwFunction{HAMMER}{HAMMER}

  \SetKwProg{Fn}{Function}{:}{\KwRet}
  \Fn{\HAMMER{$P_{\text{in}}$}}
  {
   \CommentSty{\color{blue} // Step-1: Create Hamming Spectrum}\;
      $CHS = \textrm{zeros(n/2)}$\;
      \For{$x$ $\mathbf{in}$ $P_{\text{in}}$}{
        \For{$y$ $\mathbf{in}$ $P_{\text{in}}$}{
            $d = \mathrm{Hamming\ Distance}\left({x,y}\right)$\;
            \If{ $d < \frac{n}{2} $}{
                $CHS[d] += P_{\text{in}}[y]$
            }
      }
    }
    \CommentSty{\color{blue} // Step-2: Compute Per-Distance Weights}\;
    $W = \textrm{zeros(n/2)}$\;
    \For{$i=0$ $\mathbf{to}$ $n/2$}{
        \If{ $CHS[d] > 0 $}{
            $W[d] = 1/CHS[d]$\;
        }
    }
    \CommentSty{\color{blue} // Step-3: Update the Probability Distribution}\;
    $P_{\text{out}} = \{\}$\;
    \For{$x$ $\mathbf{in}$ $P_{\text{in}}$}{
        $score = P_{\text{in}}[x]$\;
        \For{$y$ $\mathbf{in}$ $P_{\text{in}}$}{
            $d = \mathrm{Hamming\  Distance}\left({x,y}\right)$\;
            \If{ $d < \frac{n}{2}$ $\mathbf{and}$ $P_{\text{in}}[x] > P_{\text{in}}[y]$}{
                $score += W[d] \times P_{\text{in}}[y]$\;            }
        }
        $P_{\text{out}}[x] = score \times P_{\text{in}}[x]$
    }
    $P_{\text{out}} = \mathrm{normalize}\left({P_{\text{out}}}\right)$\;
    \KwRet $\mathrm{P_{out}}$\;    
  } 
  
\end{algorithm}}

\bibliographystyle{ACM-Reference-Format}
\balance
\bibliography{ref}


\begin{thebibliography}{45}


\ifx \showCODEN    \undefined \def \showCODEN     #1{\unskip}     \fi
\ifx \showDOI      \undefined \def \showDOI       #1{#1}\fi
\ifx \showISBNx    \undefined \def \showISBNx     #1{\unskip}     \fi
\ifx \showISBNxiii \undefined \def \showISBNxiii  #1{\unskip}     \fi
\ifx \showISSN     \undefined \def \showISSN      #1{\unskip}     \fi
\ifx \showLCCN     \undefined \def \showLCCN      #1{\unskip}     \fi
\ifx \shownote     \undefined \def \shownote      #1{#1}          \fi
\ifx \showarticletitle \undefined \def \showarticletitle #1{#1}   \fi
\ifx \showURL      \undefined \def \showURL       {\relax}        \fi
\providecommand\bibfield[2]{#2}
\providecommand\bibinfo[2]{#2}
\providecommand\natexlab[1]{#1}
\providecommand\showeprint[2][]{arXiv:#2}

\bibitem[\protect\citeauthoryear{AI}{AI}{2021}]%
        {sycamoredatasheet}
\bibfield{author}{\bibinfo{person}{Google~Quantum AI}.}
  \bibinfo{year}{Accessed: June 19, 2021}\natexlab{}.
\newblock \bibinfo{title}{Quantum Computer Datasheet}.
\newblock
\newblock
\newblock
\shownote{\url{https://quantumai.google/hardware/datasheet/weber.pdf}}.


\bibitem[\protect\citeauthoryear{Alam, Ash-Saki, and Ghosh}{Alam
  et~al\mbox{.}}{2020a}]%
        {alam2020accelerating}
\bibfield{author}{\bibinfo{person}{Mahabubul Alam}, \bibinfo{person}{Abdullah
  Ash-Saki}, {and} \bibinfo{person}{Swaroop Ghosh}.}
  \bibinfo{year}{2020}\natexlab{a}.
\newblock \showarticletitle{Accelerating quantum approximate optimization
  algorithm using machine learning}. In \bibinfo{booktitle}{\emph{DATE}}. IEEE,
  \bibinfo{pages}{686--689}.
\newblock
\urldef\tempurl%
\url{https://doi.org/10.23919/DATE48585.2020.9116348}
\showDOI{\tempurl}


\bibitem[\protect\citeauthoryear{Alam, Ash-Saki, and Ghosh}{Alam
  et~al\mbox{.}}{2020b}]%
        {alam2020circuit}
\bibfield{author}{\bibinfo{person}{Mahabubul Alam}, \bibinfo{person}{Abdullah
  Ash-Saki}, {and} \bibinfo{person}{Swaroop Ghosh}.}
  \bibinfo{year}{2020}\natexlab{b}.
\newblock \showarticletitle{Circuit Compilation Methodologies for Quantum
  Approximate Optimization Algorithm}. In \bibinfo{booktitle}{\emph{MICRO-53}}.
  IEEE, \bibinfo{pages}{215--228}.
\newblock
\urldef\tempurl%
\url{https://doi.org/10.1109/MICRO50266.2020.00029}
\showDOI{\tempurl}


\bibitem[\protect\citeauthoryear{Arute, Arya, Babbush, Bacon, Bardin, Barends,
  Boixo, Broughton, Buckley, Buell, et~al\mbox{.}}{Arute et~al\mbox{.}}{2020}]%
        {arute2020quantum}
\bibfield{author}{\bibinfo{person}{Frank Arute}, \bibinfo{person}{Kunal Arya},
  \bibinfo{person}{Ryan Babbush}, \bibinfo{person}{Dave Bacon},
  \bibinfo{person}{Joseph~C Bardin}, \bibinfo{person}{Rami Barends},
  \bibinfo{person}{Sergio Boixo}, \bibinfo{person}{Michael Broughton},
  \bibinfo{person}{Bob~B Buckley}, \bibinfo{person}{David~A Buell},
  {et~al\mbox{.}}} \bibinfo{year}{2020}\natexlab{}.
\newblock \showarticletitle{Quantum approximate optimization of non-planar
  graph problems on a planar superconducting processor}.
\newblock \bibinfo{journal}{\emph{arXiv preprint arXiv:2004.04197}}
  (\bibinfo{year}{2020}).
\newblock


\bibitem[\protect\citeauthoryear{Barron and Wood}{Barron and Wood}{2020}]%
        {barron2020measurement}
\bibfield{author}{\bibinfo{person}{George~S Barron} {and}
  \bibinfo{person}{Christopher~J Wood}.} \bibinfo{year}{2020}\natexlab{}.
\newblock \showarticletitle{Measurement error mitigation for variational
  quantum algorithms}.
\newblock \bibinfo{journal}{\emph{arXiv preprint arXiv:2010.08520}}
  (\bibinfo{year}{2020}).
\newblock


\bibitem[\protect\citeauthoryear{Biamonte, Wittek, Pancotti, Rebentrost, Wiebe,
  and Lloyd}{Biamonte et~al\mbox{.}}{2017}]%
        {biamonte2017quantum}
\bibfield{author}{\bibinfo{person}{Jacob Biamonte}, \bibinfo{person}{Peter
  Wittek}, \bibinfo{person}{Nicola Pancotti}, \bibinfo{person}{Patrick
  Rebentrost}, \bibinfo{person}{Nathan Wiebe}, {and} \bibinfo{person}{Seth
  Lloyd}.} \bibinfo{year}{2017}\natexlab{}.
\newblock \showarticletitle{Quantum machine learning}.
\newblock \bibinfo{journal}{\emph{Nature}} \bibinfo{volume}{549},
  \bibinfo{number}{7671} (\bibinfo{year}{2017}), \bibinfo{pages}{195--202}.
\newblock


\bibitem[\protect\citeauthoryear{Bollob{\'a}s and B{\'e}la}{Bollob{\'a}s and
  B{\'e}la}{2001}]%
        {bollobas2001random}
\bibfield{author}{\bibinfo{person}{B{\'e}la Bollob{\'a}s} {and}
  \bibinfo{person}{Bollob{\'a}s B{\'e}la}.} \bibinfo{year}{2001}\natexlab{}.
\newblock \bibinfo{booktitle}{\emph{Random graphs}}.
\newblock Number~73. \bibinfo{publisher}{Cambridge university press}.
\newblock


\bibitem[\protect\citeauthoryear{Bravyi, Sheldon, Kandala, Mckay, and
  Gambetta}{Bravyi et~al\mbox{.}}{2020}]%
        {bravyi2020mitigating}
\bibfield{author}{\bibinfo{person}{Sergey Bravyi}, \bibinfo{person}{Sarah
  Sheldon}, \bibinfo{person}{Abhinav Kandala}, \bibinfo{person}{David~C Mckay},
  {and} \bibinfo{person}{Jay~M Gambetta}.} \bibinfo{year}{2020}\natexlab{}.
\newblock \showarticletitle{Mitigating measurement errors in multi-qubit
  experiments}.
\newblock \bibinfo{journal}{\emph{arXiv preprint arXiv:2006.14044}}
  (\bibinfo{year}{2020}).
\newblock


\bibitem[\protect\citeauthoryear{Chong}{Chong}{2018}]%
        {chong2018closing}
\bibfield{author}{\bibinfo{person}{Frederic Chong}.}
  \bibinfo{year}{2018}\natexlab{}.
\newblock \showarticletitle{Closing the gap between quantum algorithms and
  hardware through software-enabled vertical integration and co-design}. In
  \bibinfo{booktitle}{\emph{APS March Meeting Abstracts}},
  Vol.~\bibinfo{volume}{2018}. \bibinfo{pages}{C39--004}.
\newblock


\bibitem[\protect\citeauthoryear{Corporation}{Corporation}{2017a}]%
        {qiskit}
\bibfield{author}{\bibinfo{person}{International Business~Machines
  Corporation}.} \bibinfo{year}{2017}\natexlab{a}.
\newblock \bibinfo{title}{{Quantum Software Development Kit for writing quantum
  computing experiments, programs, and applications}}.
\newblock
  \bibinfo{howpublished}{\url{https://github.com/QISKit/qiskit-sdk-py}}.
\newblock
\newblock
\shownote{[Online; accessed 28-AUGUST-2020]}.


\bibitem[\protect\citeauthoryear{Corporation}{Corporation}{2017b}]%
        {IBMQ}
\bibfield{author}{\bibinfo{person}{International Business~Machines
  Corporation}.} \bibinfo{year}{2017}\natexlab{b}.
\newblock \bibinfo{title}{{Universal Quantum Computer Development at IBM:}}.
\newblock
  \bibinfo{howpublished}{\url{http://research.ibm.com/ibm-q/research/}}.
\newblock
\newblock
\shownote{[Online; accessed 3-April-2017]}.


\bibitem[\protect\citeauthoryear{Cross, Bishop, Sheldon, Nation, and
  Gambetta}{Cross et~al\mbox{.}}{2019}]%
        {quantumvolume}
\bibfield{author}{\bibinfo{person}{Andrew~W Cross}, \bibinfo{person}{Lev~S
  Bishop}, \bibinfo{person}{Sarah Sheldon}, \bibinfo{person}{Paul~D Nation},
  {and} \bibinfo{person}{Jay~M Gambetta}.} \bibinfo{year}{2019}\natexlab{}.
\newblock \showarticletitle{Validating quantum computers using randomized model
  circuits}.
\newblock \bibinfo{journal}{\emph{Physical Review A}} \bibinfo{volume}{100},
  \bibinfo{number}{3} (\bibinfo{year}{2019}), \bibinfo{pages}{032328}.
\newblock


\bibitem[\protect\citeauthoryear{Das, Tannu, Dangwal, and Qureshi}{Das
  et~al\mbox{.}}{2021b}]%
        {das2021adapt}
\bibfield{author}{\bibinfo{person}{Poulami Das}, \bibinfo{person}{Swamit
  Tannu}, \bibinfo{person}{Siddharth Dangwal}, {and} \bibinfo{person}{Moinuddin
  Qureshi}.} \bibinfo{year}{2021}\natexlab{b}.
\newblock \showarticletitle{ADAPT: Mitigating Idling Errors in Qubits via
  Adaptive Dynamical Decoupling}. In \bibinfo{booktitle}{\emph{MICRO-54}}.
  \bibinfo{pages}{950--962}.
\newblock
\urldef\tempurl%
\url{https://doi.org/10.1145/3466752.3480059}
\showDOI{\tempurl}


\bibitem[\protect\citeauthoryear{Das, Tannu, and Qureshi}{Das
  et~al\mbox{.}}{2021a}]%
        {das2021jigsaw}
\bibfield{author}{\bibinfo{person}{Poulami Das}, \bibinfo{person}{Swamit
  Tannu}, {and} \bibinfo{person}{Moinuddin Qureshi}.}
  \bibinfo{year}{2021}\natexlab{a}.
\newblock \showarticletitle{JigSaw: Boosting Fidelity of NISQ Programs via
  Measurement Subsetting}. In \bibinfo{booktitle}{\emph{MICRO-54}}.
  \bibinfo{pages}{937--949}.
\newblock
\urldef\tempurl%
\url{https://doi.org/10.1145/3466752.3480044}
\showDOI{\tempurl}


\bibitem[\protect\citeauthoryear{Das, Tannu, Nair, and Qureshi}{Das
  et~al\mbox{.}}{2019}]%
        {micro3}
\bibfield{author}{\bibinfo{person}{Poulami Das}, \bibinfo{person}{Swamit~S.
  Tannu}, \bibinfo{person}{Prashant~J. Nair}, {and} \bibinfo{person}{Moinuddin
  Qureshi}.} \bibinfo{year}{2019}\natexlab{}.
\newblock \showarticletitle{A Case for Multi-Programming Quantum Computers}. In
  \bibinfo{booktitle}{\emph{MICRO-52}} (Columbus, OH, USA)
  \emph{(\bibinfo{series}{MICRO '52})}. \bibinfo{publisher}{Association for
  Computing Machinery}, \bibinfo{address}{New York, NY, USA},
  \bibinfo{pages}{291–303}.
\newblock
\showISBNx{9781450369381}
\urldef\tempurl%
\url{https://doi.org/10.1145/3352460.3358287}
\showDOI{\tempurl}


\bibitem[\protect\citeauthoryear{Farhi, Goldstone, and Gutmann}{Farhi
  et~al\mbox{.}}{2014}]%
        {qaoa}
\bibfield{author}{\bibinfo{person}{Edward Farhi}, \bibinfo{person}{Jeffrey
  Goldstone}, {and} \bibinfo{person}{Sam Gutmann}.}
  \bibinfo{year}{2014}\natexlab{}.
\newblock \showarticletitle{A quantum approximate optimization algorithm}.
\newblock \bibinfo{journal}{\emph{arXiv preprint:1411.4028}}
  (\bibinfo{year}{2014}).
\newblock


\bibitem[\protect\citeauthoryear{Funcke, Hartung, Jansen, K{\"u}hn, Stornati,
  and Wang}{Funcke et~al\mbox{.}}{2020}]%
        {funcke2020measurement}
\bibfield{author}{\bibinfo{person}{Lena Funcke}, \bibinfo{person}{Tobias
  Hartung}, \bibinfo{person}{Karl Jansen}, \bibinfo{person}{Stefan K{\"u}hn},
  \bibinfo{person}{Paolo Stornati}, {and} \bibinfo{person}{Xiaoyang Wang}.}
  \bibinfo{year}{2020}\natexlab{}.
\newblock \showarticletitle{{Measurement Error Mitigation in Quantum Computers
  Through Classical Bit-Flip Correction}}.
\newblock \bibinfo{journal}{\emph{arXiv preprint arXiv:2007.03663}}
  (\bibinfo{year}{2020}).
\newblock


\bibitem[\protect\citeauthoryear{Gokhale, Ding, Propson, Winkler, Leung, Shi,
  Schuster, Hoffmann, and Chong}{Gokhale et~al\mbox{.}}{2019}]%
        {gokhale2019partial}
\bibfield{author}{\bibinfo{person}{Pranav Gokhale}, \bibinfo{person}{Yongshan
  Ding}, \bibinfo{person}{Thomas Propson}, \bibinfo{person}{Christopher
  Winkler}, \bibinfo{person}{Nelson Leung}, \bibinfo{person}{Yunong Shi},
  \bibinfo{person}{David~I Schuster}, \bibinfo{person}{Henry Hoffmann}, {and}
  \bibinfo{person}{Frederic~T Chong}.} \bibinfo{year}{2019}\natexlab{}.
\newblock \showarticletitle{Partial Compilation of Variational Algorithms for
  Noisy Intermediate-Scale Quantum Machines}. In
  \bibinfo{booktitle}{\emph{MICRO-52}}. ACM, \bibinfo{pages}{266--278}.
\newblock
\urldef\tempurl%
\url{https://doi.org/10.1145/3352460.3358313}
\showDOI{\tempurl}


\bibitem[\protect\citeauthoryear{Gokhale, Javadi-Abhari, Earnest, Shi, and
  Chong}{Gokhale et~al\mbox{.}}{2020}]%
        {gokhale2020optimized}
\bibfield{author}{\bibinfo{person}{Pranav Gokhale}, \bibinfo{person}{Ali
  Javadi-Abhari}, \bibinfo{person}{Nathan Earnest}, \bibinfo{person}{Yunong
  Shi}, {and} \bibinfo{person}{Frederic~T Chong}.}
  \bibinfo{year}{2020}\natexlab{}.
\newblock \showarticletitle{Optimized quantum compilation for near-term
  algorithms with openpulse}. In \bibinfo{booktitle}{\emph{MICRO-53}}. IEEE,
  \bibinfo{pages}{186--200}.
\newblock


\bibitem[\protect\citeauthoryear{Harrigan, Sung, Neeley, Satzinger, Arute,
  Arya, Atalaya, Bardin, Barends, Boixo, et~al\mbox{.}}{Harrigan
  et~al\mbox{.}}{2021}]%
        {harrigan2021quantum}
\bibfield{author}{\bibinfo{person}{Matthew~P Harrigan},
  \bibinfo{person}{Kevin~J Sung}, \bibinfo{person}{Matthew Neeley},
  \bibinfo{person}{Kevin~J Satzinger}, \bibinfo{person}{Frank Arute},
  \bibinfo{person}{Kunal Arya}, \bibinfo{person}{Juan Atalaya},
  \bibinfo{person}{Joseph~C Bardin}, \bibinfo{person}{Rami Barends},
  \bibinfo{person}{Sergio Boixo}, {et~al\mbox{.}}}
  \bibinfo{year}{2021}\natexlab{}.
\newblock \showarticletitle{Quantum approximate optimization of non-planar
  graph problems on a planar superconducting processor}.
\newblock \bibinfo{journal}{\emph{Nature Physics}} \bibinfo{volume}{17},
  \bibinfo{number}{3} (\bibinfo{year}{2021}), \bibinfo{pages}{332--336}.
\newblock
\urldef\tempurl%
\url{https://doi.org/10.1038/s41567-020-01105-y}
\showDOI{\tempurl}


\bibitem[\protect\citeauthoryear{IBM}{IBM}{2019}]%
        {matrixmeasurementmitigation}
\bibfield{author}{\bibinfo{person}{IBM}.} \bibinfo{year}{2019}\natexlab{}.
\newblock \bibinfo{title}{{Measurement Error Mitigation}}.
\newblock
  \bibinfo{howpublished}{\url{https://qiskit.org/textbook/ch-quantum-hardware/measurement-error-mitigation.html}}.
\newblock
\newblock
\shownote{[Online; accessed 26-July-2020]}.


\bibitem[\protect\citeauthoryear{IBM}{IBM}{2021}]%
        {ibmq1000qubitroadmap}
\bibfield{author}{\bibinfo{person}{IBM}.} \bibinfo{year}{2021}\natexlab{}.
\newblock \bibinfo{title}{{IBM’s roadmap for scaling quantum technology}}.
\newblock
\newblock
\newblock
\shownote{\url{https://research.ibm.com/blog/ibm-quantum-roadmap}}.


\bibitem[\protect\citeauthoryear{Kwon and Bae}{Kwon and Bae}{2021}]%
        {kwon2020hybrid}
\bibfield{author}{\bibinfo{person}{Hyeokjea Kwon} {and}
  \bibinfo{person}{Joonwoo Bae}.} \bibinfo{year}{2021}\natexlab{}.
\newblock \showarticletitle{A Hybrid Quantum-Classical Approach to Mitigating
  Measurement Errors in Quantum Algorithms}.
\newblock \bibinfo{journal}{\emph{IEEE Trans. Comput.}} \bibinfo{volume}{70},
  \bibinfo{number}{9} (\bibinfo{year}{2021}), \bibinfo{pages}{1401--1411}.
\newblock
\urldef\tempurl%
\url{https://doi.org/10.1109/TC.2020.3009664}
\showDOI{\tempurl}


\bibitem[\protect\citeauthoryear{Li, Ding, and Xie}{Li et~al\mbox{.}}{2019}]%
        {li2018tackling}
\bibfield{author}{\bibinfo{person}{Gushu Li}, \bibinfo{person}{Yufei Ding},
  {and} \bibinfo{person}{Yuan Xie}.} \bibinfo{year}{2019}\natexlab{}.
\newblock \showarticletitle{Tackling the qubit mapping problem for NISQ-era
  quantum devices}. In \bibinfo{booktitle}{\emph{ASPLOS-24}}.
  \bibinfo{pages}{1001--1014}.
\newblock
\urldef\tempurl%
\url{https://doi.org/10.1145/3297858.3304023}
\showDOI{\tempurl}


\bibitem[\protect\citeauthoryear{McClean, Romero, Babbush, and
  Aspuru-Guzik}{McClean et~al\mbox{.}}{2016}]%
        {vqe}
\bibfield{author}{\bibinfo{person}{Jarrod~R McClean}, \bibinfo{person}{Jonathan
  Romero}, \bibinfo{person}{Ryan Babbush}, {and} \bibinfo{person}{Al{\'{a}}n
  Aspuru-Guzik}.} \bibinfo{year}{2016}\natexlab{}.
\newblock \showarticletitle{The theory of variational hybrid quantum-classical
  algorithms}.
\newblock \bibinfo{journal}{\emph{New Journal of Physics}}
  \bibinfo{volume}{18}, \bibinfo{number}{2} (\bibinfo{date}{feb}
  \bibinfo{year}{2016}), \bibinfo{pages}{023023}.
\newblock
\urldef\tempurl%
\url{https://doi.org/10.1088/1367-2630/18/2/023023}
\showDOI{\tempurl}


\bibitem[\protect\citeauthoryear{Murali, Baker, Abhari, Chong, and
  Martonosi}{Murali et~al\mbox{.}}{2019a}]%
        {noiseadaptive}
\bibfield{author}{\bibinfo{person}{Prakash Murali}, \bibinfo{person}{Jonathan~M
  Baker}, \bibinfo{person}{Ali~Javadi Abhari}, \bibinfo{person}{Frederic~T
  Chong}, {and} \bibinfo{person}{Margaret Martonosi}.}
  \bibinfo{year}{2019}\natexlab{a}.
\newblock \showarticletitle{Noise-Adaptive Compiler Mappings for Noisy
  Intermediate-Scale Quantum Computers}.
\newblock \bibinfo{journal}{\emph{arXiv preprint arXiv:1901.11054}}
  (\bibinfo{year}{2019}).
\newblock


\bibitem[\protect\citeauthoryear{Murali, Baker, Javadi-Abhari, Chong, and
  Martonosi}{Murali et~al\mbox{.}}{2019b}]%
        {murali2019noise}
\bibfield{author}{\bibinfo{person}{Prakash Murali}, \bibinfo{person}{Jonathan~M
  Baker}, \bibinfo{person}{Ali Javadi-Abhari}, \bibinfo{person}{Frederic~T
  Chong}, {and} \bibinfo{person}{Margaret Martonosi}.}
  \bibinfo{year}{2019}\natexlab{b}.
\newblock \showarticletitle{Noise-adaptive compiler mappings for noisy
  intermediate-scale quantum computers}. In
  \bibinfo{booktitle}{\emph{ASPLOS-24}}. \bibinfo{pages}{1015--1029}.
\newblock
\urldef\tempurl%
\url{https://doi.org/10.1145/3297858.3304075}
\showDOI{\tempurl}


\bibitem[\protect\citeauthoryear{Murali, McKay, Martonosi, and
  Javadi-Abhari}{Murali et~al\mbox{.}}{2020}]%
        {murali2020software}
\bibfield{author}{\bibinfo{person}{Prakash Murali}, \bibinfo{person}{David~C
  McKay}, \bibinfo{person}{Margaret Martonosi}, {and} \bibinfo{person}{Ali
  Javadi-Abhari}.} \bibinfo{year}{2020}\natexlab{}.
\newblock \showarticletitle{Software Mitigation of Crosstalk on Noisy
  Intermediate-Scale Quantum Computers}.
\newblock \bibinfo{journal}{\emph{arXiv preprint arXiv:2001.02826}}
  (\bibinfo{year}{2020}).
\newblock


\bibitem[\protect\citeauthoryear{Nishio, Pan, Satoh, Amano, and
  Van~Meter}{Nishio et~al\mbox{.}}{2019}]%
        {nishio}
\bibfield{author}{\bibinfo{person}{Shin Nishio}, \bibinfo{person}{Yulu Pan},
  \bibinfo{person}{Takahiko Satoh}, \bibinfo{person}{Hideharu Amano}, {and}
  \bibinfo{person}{Rodney Van~Meter}.} \bibinfo{year}{2019}\natexlab{}.
\newblock \showarticletitle{Extracting Success from IBM's 20-Qubit Machines
  Using Error-Aware Compilation}.
\newblock \bibinfo{journal}{\emph{arXiv preprint arXiv:1903.10963}}
  (\bibinfo{year}{2019}).
\newblock


\bibitem[\protect\citeauthoryear{of~Sciences~Engineering and
  Medicine}{of~Sciences~Engineering and Medicine}{2019}]%
        {NAE}
\bibfield{author}{\bibinfo{person}{National~Academies of Sciences~Engineering}
  {and} \bibinfo{person}{Medicine}.} \bibinfo{year}{2019}\natexlab{}.
\newblock \bibinfo{booktitle}{\emph{Quantum Computing: Progress and
  Prospects}}.
\newblock \bibinfo{publisher}{The National Academies Press},
  \bibinfo{address}{Washington, DC}.
\newblock
\showISBNx{978-0-309-47969-1}
\urldef\tempurl%
\url{https://doi.org/10.17226/25196}
\showDOI{\tempurl}


\bibitem[\protect\citeauthoryear{Orus, Mugel, and Lizaso}{Orus
  et~al\mbox{.}}{2019}]%
        {orus2019quantum}
\bibfield{author}{\bibinfo{person}{Roman Orus}, \bibinfo{person}{Samuel Mugel},
  {and} \bibinfo{person}{Enrique Lizaso}.} \bibinfo{year}{2019}\natexlab{}.
\newblock \showarticletitle{Quantum computing for finance: overview and
  prospects}.
\newblock \bibinfo{journal}{\emph{Reviews in Physics}} (\bibinfo{year}{2019}).
\newblock
\urldef\tempurl%
\url{https://doi.org/10.1016/j.revip.2019.100028}
\showDOI{\tempurl}


\bibitem[\protect\citeauthoryear{Patel, Li, Roy, and Tiwari}{Patel
  et~al\mbox{.}}{2020}]%
        {patel2020ureqa}
\bibfield{author}{\bibinfo{person}{Tirthak Patel}, \bibinfo{person}{Baolin Li},
  \bibinfo{person}{Rohan~Basu Roy}, {and} \bibinfo{person}{Devesh Tiwari}.}
  \bibinfo{year}{2020}\natexlab{}.
\newblock \showarticletitle{$\{$UREQA$\}$: Leveraging Operation-Aware Error
  Rates for Effective Quantum Circuit Mapping on NISQ-Era Quantum Computers}.
  In \bibinfo{booktitle}{\emph{2020 $\{$USENIX$\}$ Annual Technical Conference
  ($\{$USENIX$\}$$\{$ATC$\}$ 20)}}. \bibinfo{pages}{705--711}.
\newblock


\bibitem[\protect\citeauthoryear{Patel and Tiwari}{Patel and Tiwari}{2020a}]%
        {patel2020disq}
\bibfield{author}{\bibinfo{person}{Tirthak Patel} {and} \bibinfo{person}{Devesh
  Tiwari}.} \bibinfo{year}{2020}\natexlab{a}.
\newblock \showarticletitle{DisQ: A Novel Quantum Output State Classification
  Method on IBM Quantum Computers Using Openpulse}. In
  \bibinfo{booktitle}{\emph{ICCAD-39}}. \bibinfo{publisher}{ACM}, Article
  \bibinfo{articleno}{139}, \bibinfo{numpages}{9}~pages.
\newblock
\showISBNx{9781450380263}
\urldef\tempurl%
\url{https://doi.org/10.1145/3400302.3415619}
\showDOI{\tempurl}


\bibitem[\protect\citeauthoryear{Patel and Tiwari}{Patel and Tiwari}{2020b}]%
        {patel2020veritas}
\bibfield{author}{\bibinfo{person}{Tirthak Patel} {and} \bibinfo{person}{Devesh
  Tiwari}.} \bibinfo{year}{2020}\natexlab{b}.
\newblock \showarticletitle{VERITAS: Accurately Estimating the Correct Output
  on Noisy Intermediate-Scale Quantum Computers}. In
  \bibinfo{booktitle}{\emph{SC20}}. \bibinfo{pages}{1--16}.
\newblock
\urldef\tempurl%
\url{https://doi.org/10.1109/SC41405.2020.00019}
\showDOI{\tempurl}


\bibitem[\protect\citeauthoryear{Patel and Tiwari}{Patel and Tiwari}{2021}]%
        {patel2021qraft}
\bibfield{author}{\bibinfo{person}{Tirthak Patel} {and} \bibinfo{person}{Devesh
  Tiwari}.} \bibinfo{year}{2021}\natexlab{}.
\newblock \showarticletitle{Qraft: reverse your Quantum circuit and know the
  correct program output}. In \bibinfo{booktitle}{\emph{ASPLOS-26}}.
  \bibinfo{pages}{443--455}.
\newblock
\urldef\tempurl%
\url{https://doi.org/10.1145/3445814.3446743}
\showDOI{\tempurl}


\bibitem[\protect\citeauthoryear{Preskill}{Preskill}{2018}]%
        {preskillNISQ}
\bibfield{author}{\bibinfo{person}{John Preskill}.}
  \bibinfo{year}{2018}\natexlab{}.
\newblock \showarticletitle{Quantum Computing in the NISQ era and beyond}.
\newblock \bibinfo{journal}{\emph{arXiv preprint arXiv:1801.00862}}
  (\bibinfo{year}{2018}).
\newblock


\bibitem[\protect\citeauthoryear{Quantum and Collaborators}{Quantum and
  Collaborators}{2020}]%
        {Googlefigshare}
\bibfield{author}{\bibinfo{person}{Google~AI Quantum} {and}
  \bibinfo{person}{Collaborators}.} \bibinfo{year}{2020}\natexlab{}.
\newblock \bibinfo{title}{Sycamore QAOA experimental data}.
\newblock
  \bibinfo{howpublished}{\url{"https://figshare.com/articles/dataset/Sycamore_QAOA_experimental_data/12597590"}}.
\newblock
\newblock
\shownote{[Online; accessed 26-July-2021]}.


\bibitem[\protect\citeauthoryear{Shi, Gokhale, Murali, Baker, Duckering, Ding,
  Brown, Chamberland, Javadi-Abhari, Cross, et~al\mbox{.}}{Shi
  et~al\mbox{.}}{2020}]%
        {shi2020resource}
\bibfield{author}{\bibinfo{person}{Yunong Shi}, \bibinfo{person}{Pranav
  Gokhale}, \bibinfo{person}{Prakash Murali}, \bibinfo{person}{Jonathan~M
  Baker}, \bibinfo{person}{Casey Duckering}, \bibinfo{person}{Yongshan Ding},
  \bibinfo{person}{Natalie~C Brown}, \bibinfo{person}{Christopher Chamberland},
  \bibinfo{person}{Ali Javadi-Abhari}, \bibinfo{person}{Andrew~W Cross},
  {et~al\mbox{.}}} \bibinfo{year}{2020}\natexlab{}.
\newblock \showarticletitle{Resource-Efficient Quantum Computing by Breaking
  Abstractions}.
\newblock \bibinfo{journal}{\emph{Proc. IEEE}} (\bibinfo{year}{2020}).
\newblock


\bibitem[\protect\citeauthoryear{Shi, Leung, Gokhale, Rossi, Schuster,
  Hoffmann, and Chong}{Shi et~al\mbox{.}}{2019}]%
        {shi2019optimized}
\bibfield{author}{\bibinfo{person}{Yunong Shi}, \bibinfo{person}{Nelson Leung},
  \bibinfo{person}{Pranav Gokhale}, \bibinfo{person}{Zane Rossi},
  \bibinfo{person}{David~I Schuster}, \bibinfo{person}{Henry Hoffmann}, {and}
  \bibinfo{person}{Frederic~T Chong}.} \bibinfo{year}{2019}\natexlab{}.
\newblock \showarticletitle{Optimized compilation of aggregated instructions
  for realistic quantum computers}. In \bibinfo{booktitle}{\emph{ASPLOS-24}}.
\newblock
\urldef\tempurl%
\url{https://doi.org/10.1145/3297858.3304018}
\showDOI{\tempurl}


\bibitem[\protect\citeauthoryear{Smith, Ravi, Murali, Baker, Earnest,
  Javadi-Abhari, and Chong}{Smith et~al\mbox{.}}{2021}]%
        {smith2021error}
\bibfield{author}{\bibinfo{person}{Kaitlin~N Smith},
  \bibinfo{person}{Gokul~Subramanian Ravi}, \bibinfo{person}{Prakash Murali},
  \bibinfo{person}{Jonathan~M Baker}, \bibinfo{person}{Nathan Earnest},
  \bibinfo{person}{Ali Javadi-Abhari}, {and} \bibinfo{person}{Frederic~T
  Chong}.} \bibinfo{year}{2021}\natexlab{}.
\newblock \showarticletitle{Error Mitigation in Quantum Computers through
  Instruction Scheduling}.
\newblock \bibinfo{journal}{\emph{arXiv preprint arXiv:2105.01760}}
  (\bibinfo{year}{2021}).
\newblock


\bibitem[\protect\citeauthoryear{Tang, Tomesh, Suchara, Larson, and
  Martonosi}{Tang et~al\mbox{.}}{2021}]%
        {tang2021cutqc}
\bibfield{author}{\bibinfo{person}{Wei Tang}, \bibinfo{person}{Teague Tomesh},
  \bibinfo{person}{Martin Suchara}, \bibinfo{person}{Jeffrey Larson}, {and}
  \bibinfo{person}{Margaret Martonosi}.} \bibinfo{year}{2021}\natexlab{}.
\newblock \bibinfo{booktitle}{\emph{CutQC: Using Small Quantum Computers for
  Large Quantum Circuit Evaluations}}.
\newblock \bibinfo{publisher}{ACM}, \bibinfo{pages}{473–486}.
\newblock
\showISBNx{9781450383172}
\urldef\tempurl%
\url{https://doi.org/10.1145/3445814.3446758}
\showURL{%
\tempurl}


\bibitem[\protect\citeauthoryear{Tannu and Qureshi}{Tannu and Qureshi}{2019a}]%
        {micro1}
\bibfield{author}{\bibinfo{person}{Swamit~S. Tannu} {and}
  \bibinfo{person}{Moinuddin Qureshi}.} \bibinfo{year}{2019}\natexlab{a}.
\newblock \showarticletitle{Ensemble of Diverse Mappings: Improving Reliability
  of Quantum Computers by Orchestrating Dissimilar Mistakes}. In
  \bibinfo{booktitle}{\emph{MICRO-52}} (Columbus, OH, USA)
  \emph{(\bibinfo{series}{MICRO '52})}. \bibinfo{publisher}{ACM},
  \bibinfo{pages}{253–265}.
\newblock
\showISBNx{9781450369381}
\urldef\tempurl%
\url{https://doi.org/10.1145/3352460.3358257}
\showDOI{\tempurl}


\bibitem[\protect\citeauthoryear{Tannu and Qureshi}{Tannu and Qureshi}{2019b}]%
        {FNM}
\bibfield{author}{\bibinfo{person}{Swamit~S. Tannu} {and}
  \bibinfo{person}{Moinuddin~K. Qureshi}.} \bibinfo{year}{2019}\natexlab{b}.
\newblock \showarticletitle{Mitigating Measurement Errors in Quantum Computers
  by Exploiting State-Dependent Bias}. In \bibinfo{booktitle}{\emph{MICRO-52}}.
  \bibinfo{publisher}{ACM}, \bibinfo{pages}{279–290}.
\newblock
\showISBNx{9781450369381}
\urldef\tempurl%
\url{https://doi.org/10.1145/3352460.3358265}
\showDOI{\tempurl}


\bibitem[\protect\citeauthoryear{Tannu and Qureshi}{Tannu and Qureshi}{2019c}]%
        {tannu2019not}
\bibfield{author}{\bibinfo{person}{Swamit~S Tannu} {and}
  \bibinfo{person}{Moinuddin~K Qureshi}.} \bibinfo{year}{2019}\natexlab{c}.
\newblock \showarticletitle{Not all qubits are created equal: a case for
  variability-aware policies for NISQ-era quantum computers}. In
  \bibinfo{booktitle}{\emph{Proceedings of the Twenty-Fourth International
  Conference on Architectural Support for Programming Languages and Operating
  Systems}}. \bibinfo{pages}{987--999}.
\newblock


\bibitem[\protect\citeauthoryear{Zulehner, Paler, and Wille}{Zulehner
  et~al\mbox{.}}{2018}]%
        {zulehner2018efficient}
\bibfield{author}{\bibinfo{person}{Alwin Zulehner}, \bibinfo{person}{Alexandru
  Paler}, {and} \bibinfo{person}{Robert Wille}.}
  \bibinfo{year}{2018}\natexlab{}.
\newblock \showarticletitle{Efficient mapping of quantum circuits to the IBM QX
  architectures}. In \bibinfo{booktitle}{\emph{2018 Design, Automation Test in
  Europe Conference Exhibition (DATE)}}. \bibinfo{pages}{1135--1138}.
\newblock
\urldef\tempurl%
\url{https://doi.org/10.23919/DATE.2018.8342181}
\showDOI{\tempurl}


\end{thebibliography}


\end{document}